\documentclass[draftclsnofoot,onecolumn,12pt]{IEEEtran}
\usepackage{ifpdf}
\usepackage{graphicx}
\usepackage{amssymb}

\usepackage[tbtags]{amsmath}
\usepackage{amsmath}
\usepackage[linesnumbered,ruled,commentsnumbered,longend]{algorithm2e}

\usepackage{cite}
\usepackage{epstopdf}
\usepackage{dsfont}
\usepackage[justification=centering]{caption}
\usepackage{color}
\usepackage{etoolbox}
\usepackage{subcaption}
\usepackage{balance}

\newtheorem{remark}{Remark}

\makeatletter
\def\ScaleIfNeeded{%
\ifdim\Gin@nat@width>\linewidth \linewidth \else \Gin@nat@width \fi
} \makeatother

\usepackage{mathtools}
\DeclarePairedDelimiterX{\inp}[2]{\langle}{\rangle}{#1, #2}

\newcommand{\ub}[1]{\underbrace{#1}}

\newcommand{\n}{\nonumber \\}

\newcommand{\CN}{\mathcal{CN}}

\newcommand{\I}{\boldsymbol{I}}
\newcommand{\Norm}[1]{\left\|#1\right\|}
\newcommand{\Hermitan}[1]{#1^H}

\newcommand{\Expect}[1]{\mathbb{E}\left\{#1\right\}}

\newcommand{\Eve}{\mathsf{E}}

\newcommand{\betaE}{\beta_\Eve}

\newcommand{\h}{\boldsymbol{h}}
\newcommand{\w}{\boldsymbol{w}}
\newcommand{\z}{\boldsymbol{z}}

\newcommand{\g}{\boldsymbol{g}}
\newcommand{\e}{\boldsymbol{\varepsilon}}
\newcommand{\NN}{\boldsymbol{N}}

\begin{document}

\title{Securing Downlink Massive MIMO-NOMA Networks with Artificial Noise}

\author{Ming Zeng, Nam-Phong Nguyen, Octavia A. Dobre, and H. Vincent Poor
\thanks{This work was supported in part by the Natural Sciences and Engineering Research Council of Canada (NSERC) through its Discovery program, and in part by the U.S. National Science Foundation under Grants CCF-093970 and CCF-1513915.

M. Zeng, N.-P. Nguyen, and O. A. Dobre are with Memorial University, Canada.
(e-mail: \{mzeng, nnguyen, odobre\}.mun.ca).
N.-P. Nguyen is also with Hanoi University of Science and Technology, Vietnam (email: phong.nguyennam@hust.edu.vn).}
\thanks{H. V. Poor is with the Electrical Engineering Department, Princeton University, Princeton, NJ, USA
(e-mail: poor@princeton.edu).

All authors contributed equally to the article.}
}


\maketitle

\begin{abstract}
In this paper, we focus on securing the confidential information of massive multiple-input multiple-output (MIMO) non-orthogonal multiple access (NOMA) networks by exploiting artificial noise {(AN)}.
{\color{black}An uplink training scheme is first proposed with minimum mean squared error estimation at the base station. Based on the estimated channel state information, the base station precodes the confidential information and injects the AN.   
Following this, the ergodic secrecy rate is derived for downlink transmission.}
An asymptotic secrecy performance analysis is also carried out for a large number of transmit antennas and high transmit power at the base station, respectively, to highlight the effects of key parameters on the secrecy performance of the considered system.
Based on the derived ergodic secrecy rate, we propose the joint power allocation of the uplink training phase and downlink transmission phase to maximize the sum secrecy rates of the system.
Besides, from the perspective of security, another optimization algorithm is proposed to maximize the energy efficiency.
The results show that the combination of massive MIMO technique and AN greatly benefits NOMA networks in term of the secrecy performance.
In addition, the effects of the uplink training phase and clustering process on the secrecy performance are revealed. 
Besides, the proposed optimization algorithms are compared with other baseline algorithms through simulations, and their superiority is validated.
{\color{black}Finally, it is shown that the proposed system outperforms the conventional massive MIMO orthogonal multiple access in terms of the secrecy performance.}
\end{abstract}

\IEEEpeerreviewmaketitle


\section{Introduction}
The development of Internet-of-Things demands massive connectivity over the limited radio spectrum.
This requires the next generation wireless networks deploy new multiple access technologies with better spectral efficiency{\color{black}\cite{VWS_Wong}}.
Recently, non-orthogonal multiple access (NOMA) has been introduced as a solution for this challenge \cite{23, RA}.
Power-domain NOMA allows multiple users to share the same time-frequency resource simultaneously by using superposition coding and advanced interference cancellation techniques, such as successive interference cancellation (SIC)\cite{3,  12, Z_Wei17}.
As a result, NOMA can enhance the capacity of a network in both spatial and temporal dimensions \cite{25, 18, 6, Fair}.
However, from the security viewpoint, sharing the same time-frequency resource among users imposes secrecy challenges.

Traditionally, the security issues have been handled at the higher layers using encryption approaches.
However, the development of computing technologies and the tremendous growth in the number of wireless devices
have surfaced the vulnerability of the conventional encryption methods \cite{Yang:PhySec:2015}.
As a result, physical layer security (PLS) has been introduced as an additional protecting layer to the conventional encryption methods for securing confidential information \cite{Wyner:PhySec:75:Bell}.
The principle of PLS is to take advantage of the randomness of the wireless channels to restrain the illegitimate side from overhearing the legitimate users \cite{Shiu2011}.
The community has shown a great interest in applying PLS to NOMA networks.
In\cite{Chen2018_2}, the authors investigated the secrecy outage probability (SOP) of NOMA relay networks with two types of relay, i.e., amplify-and-forward and decode-and-forward.
The paper revealed that in the high signal-to-noise
ratio regime, the SOP of the considered NOMA relay network converges to a constant value.
In \cite{Liu2017}, the secrecy performance of a stochastic NOMA network was considered, by modelling its users' locations using stochastic geometry.
The results showed that the secrecy diversity order
of the considered system is determined by that of the user pair with a poorer channel.
In \cite{Zhang2016},
the authors derived a closed-form solution for maximizing the secrecy sum rate of the NOMA while taking the users' quality of service requirements into consideration.
In \cite{He2017}, the authors investigated a NOMA system in the presence of an external eavesdropper.
The SOP of the considered system was derived and used to optimize the decoding order, transmission rates, and allocated power.
These studies have laid the initial foundation for exploiting PLS {\color{black}in} NOMA networks.

Recently, massive multiple-input multiple-output (MIMO) has become one of the key technologies for 5G network \cite{Larsson:MassiveMIMO:2014, Marzetta:massiveMIMO:2010, HienQuocNgo:MassiveMIMO:2013}.
By deploying hundreds of antennas at the base station (BS) to serve tens of users, massive MIMO exploits the high spatial resolution and large array gain to greatly enhance the throughput, spectral efficiency, and energy efficiency (EE) \cite{Hao_access, 22, Ming_TVT19}.
Massive MIMO networks are suggested to operate in time division duplex to address pilot contamination by exploiting channel reciprocity \cite{Larsson:MassiveMIMO:2014}.
In massive MIMO networks, the BS can obtain the knowledge of the channel state information (CSI) via uplink training sequences of the users and employ this knowledge to precode the transmit data.
The combination of massive MIMO and NOMA seems to be naturally matched since it can offer a great performance enhancement for a large number of users \cite{Ma2017}.
However, there are some challenges of this combination.
Since the number of orthogonal sequences for the uplink training phase is limited,
the massive number of users has to be grouped in clusters.
In a cluster, users share the same training sequence.
As a consequence, the quality of the uplink training phase can be compromised.
Therefore, the spatial resolution is decreased, which can lead to leakage of the confidential information.
There have been several studies of PLS {\color{black}for} massive MIMO-NOMA networks.
In \cite{Chen2018},
the authors have {\color{black}investigated} the secrecy performance of a NOMA massive MIMO network in the presence of an active eavesdropper.
The inter-user interference was utilized to enhance the secrecy performance of the network.
Artificial noise (AN) has proven its effectiveness to secure the legitimate side from malicious attempts \cite{Physec:massiveMIMO:AN:Zhu2016,Nguyen2017}.
Recently, in \cite{Zhang2018},
the authors have proposed a joint alignment of multi-user constellations and AN to secure the massive MIMO-NOMA networks.
By using a water filling power allocation between the constellation and AN, the error rate of the legitimate user is eliminated with a large number of antennas at the receiver, while the error rate of the eavesdropper approaches a floor when the number of eavesdropper's antennas is large. So far, it is the only work that deploys AN in NOMA networks.
{\color{black}Therefore, {\color{black}the role of AN in massive MIMO-NOMA networks is far from being} well-understood.}

In this paper, we propose an AN-based PLS method for the massive MIMO-NOMA networks in the presence of a passive eavesdropper.
In order to secure the downlink transmission, the BS uses its knowledge of CSI to precode the confidential information and inject the AN, which is different from \cite{Chen2018}.
Besides, because of the high complexity of the uplink training phase in the massive MIMO-NOMA networks, the AN approaches in \cite{Physec:massiveMIMO:AN:Zhu2016,Nguyen2017} are not suitable.
Therefore, in this paper, the AN is injected in the null-space of the effective channels of the clusters in the downlink transmission phase.
To emphasize the role of the uplink training process on the secrecy performance of the considered system, the CSI knowledge at the BS is the result of an estimation process that is more practical than the assumption of perfect CSI {\color{black}in other existing work on PLS for} massive MIMO-NOMA networks. {\color{black}To the best of our knowledge, this is the first work using AN to secure massive MIMO-NOMA networks when taking imperfect channel estimation into account.}
The contributions of this paper can be summarized as follows:
\begin{itemize}
  \item We demonstrate a framework to analyze the secrecy performance of an AN-aided massive MIMO-NOMA network while taking the imperfect channel estimation into consideration.
      In particular, the ergodic secrecy rates for users are derived. The asymptotic expressions of the legitimate and illegitimate rates {\color{black}for a large number of antennas and high transmit power at the BS} are also obtained.
      {\color{black}Note that the AN-aided massive MIMO-OMA network is a special case of the proposed system. The analysis expressions can be applied directly with the number of users in each cluster being equal to one}.
  \item The results reveal that by using a sufficiently large number of antennas at the BS, the AN only affects the eavesdropper.
      In addition, when the transmit power at the BS is sufficiently high, the secrecy performance of {\color{black}a user} depends on the AN, the intra-cluster interference, and the channel estimation error of its cluster.
  \item In order to further exploit the interference and AN, we study the maximization of the sum ergodic secrecy rate (SE) and the maximization of the EE in terms of the ergodic secrecy rates.
      In this work, the EE is defined as the sum ergodic secrecy rate over the total transmit power, which includes both the uplink and downlink powers.
      For the SE maximization problem, we first decompose it into two sub-problems, i.e., uplink and downlink power allocation (PA), based on alternating optimization. Then, we address each sub-problem using difference of convex (DC) programming. {\color{black}The EE maximization problem} is of fractional form, and can be transformed into a series of SE maximization problems, which can be solved accordingly. Numerical results show that the proposed algorithms can significantly enhance the performance of the considered system, compared with other baseline algorithms.
\end{itemize}

The rest of this paper is organized as follows.
The system and channel models are described in Section II.
The analytical expressions for the ergodic secrecy rates of the considered system are developed in Section III.
In Section IV, the optimization problems are proposed, and the solutions are discussed in Section V.
The numerical results and discussions are presented in Section VI.
Finally, we conclude the paper in Section VII.
{\color{black}\subsubsection*{Notations}
Superscript $(\cdot)^H$ stands for the conjugate transpose.
The expectation operation and Frobenius norm are denoted by $\Expect{\cdot}$ and $\Norm{\cdot}$, respectively.
$\I_{N_t}$ denotes the $N_t$-dimensional identity matrix. $\CN(\mu,\sigma^2)$ indicates complex normal distribution with $\mu$ mean and $\sigma^2$ variance.
}

\section{System and Channel Models}
\label{systemmodel}
\begin{figure}
\centering
\includegraphics[width=0.4\textwidth]{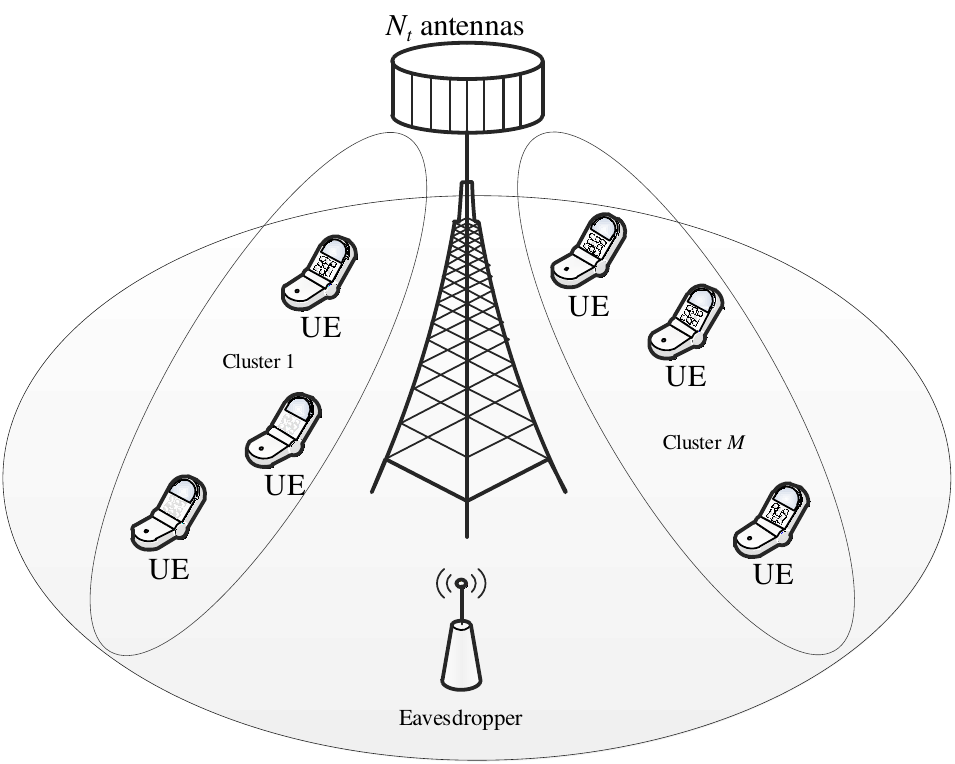}
\caption{{System model.}}
\label{systemmodel-2}
\end{figure}
{\color{black}As shown in Fig.~\ref{systemmodel-2}}, we consider the downlink transmission in a massive MIMO-NOMA system, {\color{black}which} includes one $N_t$-antenna BS, multiple single-antenna end users (UEs) that are grouped into $M$ clusters with $K_m$ users, $m=\{1,..,M\}$, in the $m$-th cluster, and one passive single-antenna eavesdropper.
Before performing the downlink transmission, the BS needs the network's CSI to precode the information and inject the AN.
Besides, the users also require knowledge of the precoding to decode the confidential information.
Therefore, the BS and users exchange CSI and precoding knowledge in the training phases.
\subsection{Training Phases}
\subsubsection{Uplink training}
During one coherence interval duration of $T$ samples, the users simultaneously send training sequences to the BS.
Users in the same cluster {\color{black}employ} the same training sequence.
In order to prevent {\color{black}the} training sequence of each cluster from interfering {\color{black}with} each other, all clusters are assigned mutually orthogonal training sequences of length $\tau$ samples, where $T\geq\tau\geq M$.
The $j$-th cluster training sequence is denoted by a $\tau \times 1$ vector $\boldsymbol{\Phi}_j$, where
$\Hermitan{\boldsymbol{\Phi}}_j\boldsymbol{\Phi}_i=0,\forall i\neq j$, $\Hermitan{\boldsymbol{\Phi}}_j\boldsymbol{\Phi}_j=1$. 
The received training signal at the BS is
\begin{align}\label{BS_uplink_training}
  \boldsymbol{Y} = \sum_{m=1}^{M} \sum_{k=1}^{K_m}
                    \sqrt{P_{m,k}\beta_{m,k}\tau}
                    \h_{m,k}
                    \Hermitan{\boldsymbol{\Phi}}_m
                    +
                    \NN,
\end{align}
where $P_{m,k}$ is the transmit power of the $k$-th UE of the $m$-th cluster, $\beta_{m,k}$ is the large-scale fading,
$\h_{m,k}$ is the small-scale fading, $\h_{m,k}\sim \CN(0,\I_{N_t})$, and the elements of $\NN\sim\CN(0,1)$ {\color{black}represent} the additive white Gaussian noise (AWGN).
Since $\boldsymbol{\Phi}_m$ is known at the BS, the BS pre-processes the received signal as follows:
\begin{align}\label{BS_estimation_process}
  \underbrace{\boldsymbol{Y} \boldsymbol{\Phi}_m}_{\tilde{\boldsymbol{y}}_m}
  &= \sum_{k=1}^{K_m}
  \sqrt{P_{m,k}\beta_{m,k}\tau}
  \h_{m,k}
  +
  \underbrace{\NN\boldsymbol{\Phi}_m}_{\tilde{\boldsymbol{n}}_m}
  \n
  &= \sqrt{\sum_{k=1}^{K_m}P_{m,k}\beta_{m,k}\tau} \h_m + \tilde{\boldsymbol{n}}_m
  ,
\end{align}
where $\h_m = \frac{\sum_{k=1}^{K_m}
  \sqrt{P_{m,k}\beta_{m,k}\tau}
  \h_{m,k}}{\sqrt{\sum_{k=1}^{K_m}P_{m,k}\beta_{m,k}\tau}}$ is the effective channel for the $m$-th cluster.

The BS uses the minimum mean squared error (MMSE) technique to estimate $\h_m$.\footnote{{\color{black}The use of MMSE has been widely addopted in massive MIMO system \cite{Marzetta:massiveMIMO:2010,HienQuocNgo:MassiveMIMO:2013}.}} 
The estimate of $\h_m$ is \cite{Chen2018}
\begin{align}\label{BS_MMSE}
  \hat{\h}_m = \frac{\sqrt{\sum_{k=1}^{K_m}P_{m,k}\beta_{m,k}\tau}}{1+\sum_{k=1}^{K_m}P_{m,k}\beta_{m,k}\tau} \tilde{\boldsymbol{y}}_m.
\end{align}

The relation between $\h_{m,k}$ and $\hat{\h}_{m}$ is
\begin{align}\label{estimation_err}
  \h_{m,k} = \sqrt{\rho_{m,k}}\hat{\h}_m + \sqrt{1-\rho_{m,k}}\e_{m,k},
\end{align}
where $\e_{m,k}\sim \CN(0,\boldsymbol{I}_{N_t})$ is the error vector, which is independent of $\hat{\h}_m$. Besides, $\rho_{m,k} = \frac{P_{m,k}\beta_{m,k}\tau}{1+\sum_{i=1}^{K_m}P_{m,i}\beta_{m,i}\tau}$ \cite{Chen2018}.
\begin{remark}
  {\color{black}For each cluster, the error of the estimation process depends on the uplink transmit power of each user, the number of users in a cluster, the large-scale fading, and the length of the training sequences.
  This error can be reduced by decreasing the number of users in a cluster.
  However, this leads to an increase in the number of clusters, and further yields more orthogonal training sequences, which are limited in certain cases, e.g., crowded stadium, busy city center, etc.}
\end{remark}

After the estimation process, the BS uses the estimates of the cluster's effective channels to precode.
In this paper, we assume that the BS employs the maximal ratio transmission (MRT) precoder, which is simple and nearly optimal in massive MIMO networks \cite{HienQuocNgo:MassiveMIMO:2013}.
The precoder is defined as
\begin{align}
\label{MRT}
  \w_m &= \frac{\hat{\h}_m}{\Norm{\hat{\h}_m}}.
\end{align}
\subsubsection{Downlink training}
The downlink training phase is similar to the uplink training phase, except that the BS uses the obtained precoder to beam the downlink pilots to the clusters.
Since the downlink pilots are known at the users, these users can estimate accurately their effective channel gains, i.e., $|\sqrt{\beta_{m,k}}\Hermitan{\h_{m,k}}\w_m|^2$.
We assume that the estimation process at users is perfect.\footnote{{\color{black}This assumption is reasonable since it has been proven that at a sufficiently high transmit power, the error of the channel estimation process at the receiver is sufficiently small and can be neglected \cite{Nguyen2017}.}}
{\color{black}Without loss of generality, the users' effective channel gains of the $m$-th cluster are ordered as follows: 
\begin{equation}
|\sqrt{\beta_{m,1}}\Hermitan{\h_{m,1}}\w_m|^2 \geq \cdots \geq |\sqrt{\beta_{m,K_m}}\Hermitan{\h_{m,K_m}}\w_m|^2.
\end{equation}
}
During this phase, the eavesdropper also obtains its effective channel gain, i.e., $|\sqrt{\betaE}\Hermitan{\g}\w_m|^2$, where $\betaE$ is the large-scale fading and $\g$ is the small-scale fading {\color{black}vector} corresponding to the eavesdropper.

\subsection{NOMA Downlink Transmission}
In order to perform NOMA downlink transmission, the BS conducts superposition coding for each cluster.
The superposition coding for the $m$-th cluster is as follows:
\begin{align}
  x_m &= \sum_{k=1}^{K_m}\sqrt{Q_{m,k}} s_{m,k},
\end{align}
where $Q_{m,k}$ is the transmit power allocated {\color{black}to} UE$_{m,k}$, and $s_{m,k}$ is the corresponding transmitted signal, satisfying $\Expect{|s_{m,k}|^2}=1$.
For securing the confidential information, the BS injects AN into the transmitted signals.
The BS combines all cluster signals as follows:
\begin{align}
  \boldsymbol{x} = \sum_{m=1}^{M} (\w_m x_m + \sqrt{Q_{m,0}}\z_m \lambda_m) ,
\end{align}
where $\w_m$ and $\z_m$ are the precoding vector and AN vector for the $m$-th cluster, respectively, $\Hermitan{\hat{\h}_m}\z_m =0$, $\Norm{\z_m}^2=1$; $Q_{m,0}$ is the power allocated for the AN and $\lambda_m$ is the AN signal of the $m$-th cluster, $\Expect{|\lambda_m|^2}=1$.

The received signal at the UE$_{m,k}$ is
\begin{align}\label{UE_received_signal}
  & y_{m,k} = \ub{\sqrt{\beta_{m,k}}\Hermitan{\h_{m,k}}\sqrt{Q_{m,k}}\w_m s_{m,k}}_{\text{Desired signal}}
  \n
  &+
  \ub{\sqrt{\beta_{m,k}} \Hermitan{\h_{m,k}}\left(\sum_{i=1,i\neq k}^{K_m} \sqrt{Q_{m,i}}\w_m s_{m,i} + \sqrt{Q_{m,0}}\z_m \lambda_{m}\right)}_{\text{Intra-cluster interference and AN}}
  \n
  &+
  \ub{\sqrt{\beta_{m,k}}\Hermitan{\h_{m,k}}\sum_{j=1,j\neq m}^{M}\left(\sum_{i=1}^{K_j}\sqrt{Q_{j,i}}\w_j s_{j,i} +\sqrt{Q_{j,0}}\z_j\lambda_{j}\right)  }_{\text{Inter-cluster interference and AN}}
  \n
  &+ n_{m,k},
\end{align}
where $n_{m,k}\sim \CN(0,1)$ is the AWGN at UE$_{m,k}$.

The eavesdropper tries to intercept the confidential information of UE$_{m,k}$. The received signal at the eavesdropper is 
\begin{align}\label{UE_received_signal}
  &y_{m,k}^e = \ub{\sqrt{\betaE}\Hermitan{\g}\sqrt{Q_{m,k}}\w_m s_{m,k}}_{\text{Desired signal}}
  \n
  &+
  \ub{\sqrt{\betaE} \Hermitan{\g} \left(\sum_{i=1,i\neq k}^{K_m} \sqrt{Q_{m,i}}\w_m s_{m,i} + \sqrt{Q_{m,0}}\z_m \lambda_{m}\right)}_{\text{Intra-cluster interference and AN}}
  \n
  &+
  \ub{\sqrt{\betaE} \Hermitan{\g}\sum_{j=1,j\neq m}^{M}\left(\sum_{i=1}^{K_j}\sqrt{Q_{j,i}}\w_j s_{j,i} +\sqrt{Q_{j,0}}\z_j\lambda_{j}\right)}_{\text{Inter-cluster interference and AN}} \n
  &+ n_{e},
\end{align}
where $n_{e}\sim\CN(0,1)$ is the AWGN at the eavesdropper.

\section{Secrecy Performance Analysis}
In this section, we {\color{black}derive} the ergodic secrecy rate of UE$_{m,k}$ from its ergodic legitimate rate and its corresponding ergodic eavesdropping rate.
\subsection{Ergodic Secrecy Rate}
The ergodic secrecy rate of UE$_{m,k}$ is
\begin{align}\label{secrate_approx}
  R^{sec}_{m,k} &= \Expect{[R_{m,k}-R^{e}_{m,k}]^+}
  \n 
  &\approx \left[\Expect{R_{m,k}}-\Expect{R^{e}_{m,k}} \right]^+,
\end{align}
where $[x]^+ = \max(x,0)$.
This approximation is reasonable in massive MIMO systems owing to the channel hardening property \cite{Nguyen:massiveMIMO:act_Eve:2017}.
The achievable rate of UE$_{m,k}$ is
\begin{align}
  \bar{R}_{m,k} = \Expect{R_{m,k}} \approx \left(1-\frac{\tau}{T}\right) \log_2(1+\bar{\gamma}_{m,k}),
\end{align}
where {\color{black}$\Expect{\cdot}$ denotes the expectation operator and}
$\bar{\gamma}_{m,k} = \frac{\kappa_{m,k}}{\sum\limits_{t=1}^{3}\Im_{m,k,t}+1}$,
with
%
%
%
%
%
%
\begin{align}
& \kappa_{m,k}  \n
&= \left|\Expect{ \sqrt{Q_{m,k}\beta_{m,k}} \Hermitan{\h_{m,k}}\w_m }\right|^2
\n
&= Q_{m,k}\beta_{m,k} \left|\Expect{ (\sqrt{\rho_{m,k}}\Hermitan{\hat{\h}}_m\w_m + \sqrt{1-\rho_{m,k}}\Hermitan{\e}_{m,k}\w_m) }\right|^2
\n
&\mathop=^{(a)} Q_{m,k}\beta_{m,k} \rho_{m,k}  \left|\Expect{\Norm{\hat{\h}_m}}\right|^2
\\
&\mathop=^{(b)} Q_{m,k}\beta_{m,k} \rho_{m,k}  \frac{\Gamma^2\left(N_t + \frac{1}{2}\right)}{\Gamma^2(N_t)}
\n
&\mathop\approx^{(c)} Q_{m,k}\beta_{m,k} \rho_{m,k} N_t, \nonumber
\end{align}
{\color{black}where step $(a)$ holds true because $\Expect{\Hermitan{\e}_{m,k}\w_m} = \Expect{\Hermitan{\e}_{m,k}}\Expect{\w_m}=0$, $\Gamma(\cdot)$ is the gamma function,
step $(b)$ is based on the fact that $\Norm{\hat{\h}_m}$ has a scaled Chi distribution with $2N_t$ degrees of freedom by a factor of $\frac{1}{\sqrt{2}}$ \cite{Nguyen2017}. Therefore, $\Expect{\Norm{\hat{\h}_m}} = \frac{\Gamma\left(N_t + \frac{1}{2}\right)}{\Gamma(N_t)}$, 
and step $(c)$ is obtained by using the approximation $\frac{\Gamma^2\left(N_t + \frac{1}{2}\right)}{\Gamma^2(N_t)} \mathop\to\limits^{N_t\to \infty} N_t$ \cite{Gradshtein2007}.
}

Further, $\Im_{m,k,i}$ for $i=\{1,2,3\}$ in the expression of $\bar{\gamma}_{m,k}$ are given {\color{black}in \eqref{m_k_1}, \eqref{m_k_2} and \eqref{m_k_3}, respectively on the top of the next page. Note that step $(a)$ in \eqref{m_k_2} is obtained because $\Hermitan{\hat{\h}_{m}}\z_m =0$ and ${\e}_{m,k}$ is independent of $\z_m$. 

It can be seen that $\Im_{m,k,1}$ denotes the desired signal leakage due to the imperfect uplink channel estimation, while $\Im_{m,k,2}$ represents the intra-cluster interference {\color{black}after SIC} and the AN leakage. In addition, $\Im_{m,k,3}$ expresses the inter-cluster interference and AN.}
\begin{figure*}[!t]
\normalsize
\begin{align} \label{m_k_1}
\Im_{m,k,1} &= Q_{m,k}\beta_{m,k}\left(\Expect{|\Hermitan{\h_{m,k}}\w_m|^2}  - \left(\Expect{\Hermitan{\h_{m,k}}\w_m }\right)^2 \right)
\n
&= Q_{m,k}\beta_{m,k}
\left(
\Expect{\left|\sqrt{\rho_{m,k}}\Hermitan{\hat{\h}}_m\w_m + \sqrt{1-\rho_{m,k}}\Hermitan{\e}_{m,k}\w_m\right|^2}  - \left(\Expect{\Hermitan{\h_{m,k}}\w_m }\right)^2
\right)
\n
&= Q_{m,k}\beta_{m,k}
\left(
\rho_{m,k}\Expect{\left|\Hermitan{\hat{\h}}_m\w_m\right|^2}
+ (1-\rho_{m,k})\Expect{\left|\Hermitan{\e}_{m,k}\w_m\right|^2}
- \left(\Expect{\Hermitan{\h_{m,k}}\w_m }\right)^2
\right)
\n
&= Q_{m,k}\beta_{m,k}
\left(
\rho_{m,k} N_t + 1 -\rho_{m,k} -\rho_{m,k}\frac{\Gamma^2\left(N_t + \frac{1}{2}\right)}{\Gamma^2(N_t)}
\right)
\n
&= Q_{m,k}\beta_{m,k}(1 -\rho_{m,k}),
\end{align}

\begin{align} \label{m_k_2}
 \Im_{m,k,2} &= \Expect{\beta_{m,k} \left(\sum\limits_{i=1}^{k-1} Q_{m,i}|\Hermitan{\h_{m,k}}\w_m|^2 + Q_{m,0}|\Hermitan{\h_{m,k}}\z_m|^2 \right)}
\n
&= \beta_{m,k}\left(\sum\limits_{i=1}^{k-1} Q_{m,i}\Expect{|\Hermitan{\h_{m,k}}\w_m|^2} + Q_{m,0}\Expect{|\Hermitan{\h_{m,k}}\z_m|^2}\right)
\n
&\mathop=^{(a)} \beta_{m,k}\left[\sum\limits_{i=1}^{k-1} Q_{m,i} (\rho_{m,k} N_t +1 -\rho_{m,k}) + Q_{m,0}(1-\rho_{m,k})\right],
\end{align}

\begin{align} \label{m_k_3}
\Im_{m,k,3} &=\Expect{\beta_{m,k}\sum\limits_{j=1,j\neq m}^{M}\left( \sum\limits_{i=1}^{K_j} Q_{j,i} |\Hermitan{\h_{m,k}}\w_j|^2 + Q_{j,0}|\Hermitan{\h_{m,k}}\z_j|^2\right)}
\n
&= \beta_{m,k}\sum\limits_{j=1,j\neq m}^{M}\sum\limits_{i=0}^{K_j} Q_{j,i}.
\end{align}
\end{figure*}

{\color{black}
\begin{remark}
Note that perfect SIC is assumed to obtain $\Im_{m,k,2}$. That is, the $k$-th user first decodes and subtracts the interfering signals from the $K_m$-th to the $(k+ 1)$-th user in sequence, and then demodulates its desired signal $s_{m,k}$. In other words, the residual intra-cluster
interference is only from the users with stronger channel
gains, i.e., the first user to the $(k-1)$-th user. In practice, owing to channel estimation error, hardware limitation, low signal quality, and so on, the decoding error of the weak interfering signal may occur. Consequently, there exists residual interference from the weak users after SIC, namely imperfect SIC. This residual interference is similar to the intra-cluster interference. As shown in \cite{H_T17, H_Sun16, X_Chen18}, the residual interference can be modeled as a linear function of the power of the interfering signal, and the coefficient of imperfect SIC can be obtained through long-term measurements. As a result, the ergodic secrecy rate in the presence of imperfect SIC can be directly derived by adding the term of residual interference in $\Im_{m,k,2}$.
\end{remark}
}

The ergodic eavesdropping rate corresponding to UE$_{m,k}$ is
\begin{align}
  \bar{R}^{e}_{m,k} = \Expect{R^{e}_{m,k}} \approx  \left(1-\frac{\tau}{T}\right) \log_2(1+\bar{\gamma}_{m,k}^{e}),
\end{align}
where $\bar{\gamma}_{m,k}^{e} = \frac{\kappa_{m,k}^e}{\sum\limits_{t=1}^{2}\Im_{m,k,t}^e +1}$, with

\begin{align*}
\kappa_{m,k}^e &= Q_{m,k}\betaE \Expect{|\Hermitan{\g}\w_m|^2} = Q_{m,k}\betaE, \\ 
\Im_{m,k,1}^e & = \sum\limits_{i=1,i\neq k}^{K_m} Q_{m,i}\betaE \Expect{|\Hermitan{\g}\w_m|^2} \\
&+ Q_{m,0}\betaE \Expect{\left|\Hermitan{\g}\z_m \right|^2} \\
& = \betaE \sum\limits_{i=0,i\neq k}^{K_m} Q_{m,i}, \\
\Im_{m,k,2}^e &= \betaE\sum\limits_{j=1,j\neq m}^{M} \left(\sum\limits_{i=1}^{K_j} Q_{j,i} \Expect{|\Hermitan{\g}\w_j|^2}\right.\\ 
&+ \left. Q_{j,0}\Expect{\left|\Hermitan{\g}\z_m \right|^2} \right) \\
& = \betaE \sum\limits_{j=1,j\neq m}^{M}  \sum\limits_{i=0}^{K_j} Q_{j,i}.
\end{align*}

Therefore, $\bar{R}^{e}_{m,k}$ can be simplified as \eqref{Eavesdropping_rate} on the top of the next page.\footnote{It is possible to extend this work to the case of multiple eavesdroppers or multi-antenna eavesdropper since \eqref{Eavesdropping_rate} can be applied to each eavesdropper or each antenna of a {multi-antenna} eavesdropper. The secrecy performance in these cases is determined by the strongest eavesdropper or the strongest eavesdropping antenna.}  
\begin{figure*}[!t]
\normalsize
\begin{align}\label{Eavesdropping_rate}
  \bar{R}^{e}_{m,k} = \left(1-\frac{\tau}{T}\right) \log_2 \left(
  1 +
  \frac{Q_{m,k}\betaE}
  {\betaE \sum_{i=0,i\neq k}^{K_m} Q_{m,i}
  + \betaE\sum_{j=1,j\neq m}^{M} \sum_{i=0}^{K_j} Q_{j,i} + 1}
  \right).
\end{align}
\end{figure*}


By comparing the intra-cluster interference terms in $ \bar{R}_{m,k}$ and $\bar{R}^{e}_{m,k} $, i.e., $\Im_{m,k,2}$ and $ \Im_{m,k,1}^e$, we can observe that the intra-cluster interference has less impact on the legitimate users owing to SIC. This helps to achieve a higher secrecy rate.

\subsection{Asymptotic Secrecy Performance}
In this subsection, {\color{black}increasing the number of antennas and the transmit power at the BS are respectively studied to reveal insights into the considered system}.
\subsubsection{Large Number of Antennas at the BS}
We first investigate the impact of a large number of antennas at the BS on the secrecy performance.
From \eqref{Eavesdropping_rate}, we can observe that the eavesdropping rate is independent {\color{black}of} the number of antennas at the BS.
When this number is large, the legitimate rate is expressed as
\begin{align}
   \bar{R}_{m,k}\mathop =^{N_t\to\infty} \left(1-\frac{\tau}{T}\right) \log_2
   \left(
   1+
   \frac{Q_{m,k}}
   {\sum\limits_{i=1}^{k-1} Q_{m,i}
   }
   \right).
\end{align}
\begin{remark}
When the number of antennas at the BS is sufficiently large, the secrecy rate converges to a constant value.
At the legitimate side, the effect of imperfect CSI, fading, inter-cluster interference, and AN leakage is negligible because of channel hardening.
The legitimate rate depends only on the intra-cluster transmit powers.
Meanwhile, the eavesdropping rate suffers from noise, interferences, and fading.
Obviously, by using AN, the secrecy performance can be guaranteed in this scenario.
\end{remark}

\subsubsection{High Transmit Power at the BS}
In order to reveal the impact of the transmit power at the BS, the transmit power for each user is set proportional to the maximum transmit power of the BS, i.e., $Q_{m,k} = \sigma_{m,k}Q_{\max}$, where $Q_{\max}$ is the maximum transmit power at the BS and $\sum\limits_{m=1}^{M}\sum\limits_{k=1}^{K_m}\sigma_{m,k}=1$.
When $Q_{\max}$ is large, the legitimate rate and the eavesdropping rate are respectively approximated as {\color{black}\eqref{m_k_appr} and \eqref{m_k_e_appr} on the top of the next page.} 
\begin{figure*}[!t]
\normalsize
\begin{align} \label{m_k_appr}
  &\bar{R}_{m,k} \mathop=^{Q_{\max}\to\infty} \left(1-\frac{\tau}{T}\right)
  \n
  &\times\log_2
  \left(1+
    \frac{\sigma_{m,k} \rho_{m,k} N_t}
    { \sigma_{m,k}(1 -\rho_{m,k})
    + \left[\sum\limits_{i=1}^{k-1} \sigma_{m,i} (\rho_{m,k} N_t +1 -\rho_{m,k}) + \sigma_{m,0}(1-\rho_{m,k})\right]
    + \sum\limits_{j=1,j\neq m}^{M}\sum\limits_{i=0}^{K_j} \sigma_{i,j}
    }
  \right),
\end{align}

\begin{align} \label{m_k_e_appr}
  \bar{R}^{e}_{m,k} \mathop=^{Q_{\max}\to\infty} \left(1-\frac{\tau}{T}\right)
   \log_2 \left(
  1 +
  \frac{\sigma_{m,k}}
  {\sum_{i=0,i\neq k}^{K_m} \sigma_{m,i}
  +\sum_{j=1,j\neq m}^{M} \sum_{i=0}^{K_j} \sigma_{i,j}
  }
  \right).
\end{align}
\end{figure*}
\begin{remark}
  When the transmit power at the BS is high, we can observe that:
  \begin{itemize}
    \item The secrecy rate converges to a constant value. This value is independent of fading and the maximum transmit power.
    \item The legitimate rate and the eavesdropping rate suffer from the same amount of inter-cluster interference and inter-cluster AN.
    In other words, the secrecy rate is independent of the inter-cluster interference and inter-cluster AN.
    \item The eavesdropper is affected by the AN more heavily than the legitimate user.
    This effect depends on the uplink training process.
    Recalling Remark~1, we can conclude that the secrecy performance depends on the number of available orthogonal pilots.
  \end{itemize}
\end{remark}

\section{Optimization Problems}
{\color{black}In this section, we consider the optimization of the uplink and downlink {\color{black}PA} to fully exploit the potential of the proposed secure massive MIMO-NOMA network. Two system level criteria are respectively considered, i.e., the SE maximization and the EE maximization.}

\subsection{SE Maximization}
First, we aim to maximize the SE for the considered system, which is formulated as
\begin{IEEEeqnarray*}{clr}\label{eq:OB}
\displaystyle {\underset{\bf{P}, \bf{Q}}{\rm{max}} }  &~ \sum_{m=1}^M \sum_{k=1}^{K_m} R^{sec}_{m,k} \IEEEyesnumber \IEEEyessubnumber* \\
\text{s.t.} & 0 \leq P_{m,k} \leq P_{m,k}^{\text{max}}, m\in\{1, \cdots, M\},\\
&~k\in\{1, \cdots, K_m\},  \n
& Q_{m,k} \geq 0,  m\in\{1, \cdots, M\}, k\in\{0, \cdots, K_m\}, \\
& \sum_{m=1}^M \sum_{k=0}^{K_m} Q_{m,k} \leq Q_{\rm{max}},
\end{IEEEeqnarray*}
where ${\bf{P}} \in \mathcal{R}^{M \times K_m}$ and $ {\bf{Q}} \in \mathcal{R}^{M \times (K_m+1)}$ denote the matrix for the uplink and downlink power, respectively. {\color{black}Equations} (\ref{eq:OB}b) and (\ref{eq:OB}d) {\color{black}represent} the maximum transmit power constraint for each user in uplink and the total power constraint in downlink, respectively. Note that there exists a {\color{black}one-to-one} mapping between $P_{m,k}$ and $\rho_{m,k}$.

\subsection{EE {\color{black}Maximization}}
We also consider {\color{black}maximization of EE,} defined as the sum ergodic secrecy rate over the total transmit power, which includes both the uplink and downlink power \cite{EE, Ming_ICC18, Hao_IoT18}. Moreover, for uplink and downlink power, both fixed circuit power and dynamic transmit power are considered \cite{Ming_GC18, Ming_IoT19}. We denote the overall circuit power of the system as $P_f$. Then, the EE is given as
\begin{equation}
\eta_{\rm{EE}} =\frac{\sum_{m=1}^M \sum_{k=1}^{K_m} R^{sec}_{m,k}}{\sum_{m=1}^M \sum_{k=1}^{K_m} P_{m,k}+ \sum_{m=1}^M \sum_{k=0}^{K_m} Q_{m,k} +P_f} .
\end{equation}

Accordingly, the EE optimization problem can be expressed as
\begin{equation} \label{eq:EE}
{\underset{\bf{P}, \bf{Q}}{\rm{max}} } ~ \eta_{\rm{EE}}, ~{\rm{s.t.}}~(\ref{eq:OB}\rm{b})-(\ref{eq:OB}\rm{d}).
\end{equation}

\section{Proposed Solutions}
\subsection{SE Maximization}
Problem \eqref{eq:OB} is clearly non-convex, owing to the non-convex objective function. Moreover, it can be seen that the uplink power ${\bf{P}}$ and downlink power ${\bf{Q}}$ are coupled in the objective function. This coupling makes \eqref{eq:OB} difficult to handle. To address it, we propose to decompose the original problem into the following two sub-problems:

\subsubsection{Uplink Power Allocation for Channel Estimation}
For this sub-problem, we assume that the downlink power is appropriately allocated to the users and the AN, i.e., ${\bf{Q}}$ is known and given. Then, the original problem can be simplified as
\begin{equation} \label{eq:OB 1}
{\underset{\bf{P} }{\rm{max}} }  ~ \sum_{m=1}^M \sum_{k=1}^{K_m} R^{sec}_{m,k}, ~{\rm{s.t.}}~ (\ref{eq:OB}\rm{b}).
\end{equation}


\subsubsection{Downlink Power Allocation for Data Transmission}
Likewise, here we assume that the uplink power is appropriately allocated to the users, i.e., ${\bf{P}}$ is known and given. Then, the original problem is re-expressed as
\begin{equation} \label{eq:OB 2}
{\underset{\bf{P} }{\rm{max}} }  ~ \sum_{m=1}^M \sum_{k=1}^{K_m} R^{sec}_{m,k}, ~{\rm{s.t.}}~ (\ref{eq:OB}\rm{c}), (\ref{eq:OB}\rm{d}).
\end{equation}


For sub-problem (1), since ${\bf{Q}}$ is given, it can be seen that $\bar{R}^{e}_{m,k}$ is a constant. Then, we only need to consider $\bar{R}_{m,k}$. After some mathematical manipulations, $\bar{R}_{m,k} $ can be expressed as
\begin{align}
 & \bar{R}_{m,k}  = (1-\frac{\tau}{T})  \log_2\left(1+\frac{\kappa_{m,k} }{\sum\limits_{t=1}^{3}\Im_{m,k,t} +1}\right) \\
  &=(1-\frac{\tau}{T})  
\n
&\times  
  \log_2\left(1+\frac{a_1 \beta_{m,k} \tau P_{m,k}}{a_2 \beta_{m,k} \tau P_{m,k} + a_3 \tau \sum_{i=1}^{K_m}\beta_{m,i} P_{m,i} +a_3 }\right), \nonumber
\end{align}
where $a_1=Q_{m,k}\beta_{m,k} N_t$, $a_2=\beta_{m,k}[(N_t-1)\sum_{i=1}^{k-1}Q_{m,i}-Q_{m,0}-Q_{m,k} ] $, $a_3=\beta_{m,k} \sum_{i=0}^{k}Q_{m,i}+ \beta_{m,k} \sum_{j \neq m} \sum_{i=0}^{K_j}Q_{j,i}+1$.


On this basis, we further transform $f= \sum_{m=1}^M \sum_{k=1}^{K_m} \bar{R}_{m,k} $ as {\color{black}\eqref{f} on the top of the next page.}
\begin{figure*}[!t]
\normalsize
\begin{align} \label{f}
  f
  &=(1-\frac{\tau}{T})   \sum_{m=1}^M \sum_{k=1}^{K_m} \ub{ \log_2\left({(a_1+a_2) \beta_{m,k} \tau P_{m,k} + a_3 \tau \sum_{i=1}^{K_m}\beta_{m,i} P_{m,i} +a_3 }\right) }_{f_1(\mathbf{P})}  \n
  &~-(1-\frac{\tau}{T})   \sum_{m=1}^M \sum_{k=1}^{K_m} \ub{ \log_2\left({a_2 \beta_{m,k} \tau P_{m,k} + a_3 \tau \sum_{i=1}^{K_m}\beta_{m,i} P_{m,i} +a_3 }\right) }_{f_2(\mathbf{P})}.
\end{align}
\end{figure*}

{\color{black}Note that $(1-\frac{\tau}{T})$ is a constant, which does not affect the solution and can be removed.} Then, \eqref{eq:OB 1} can be re-expressed as
\begin{equation} \label{dc_programming}
\underset{\bf{P}}{\rm{max}}~~\sum_{m=1}^M \sum_{k=1}^{K_m} f_1(\mathbf{P})-f_2(\mathbf{P}), ~{\rm{s.t.}}~ (\ref{eq:OB}\rm{b}),
\end{equation}
where both functions $ f_1(\mathbf{P})$ and $f_2(\mathbf{P})$ are concave. Thus, the objective $\sum_{m=1}^M \sum_{k=1}^{K_m} f_1(\mathbf{P})-f_2(\mathbf{P})$ is a DC function. The gradient of $f_2$ at $P_{j,i}, \forall j \in \{1, \cdots, M\}, i \in \{1, \cdots, K_j\}$ is given by
\begin{align*}
&\nabla f_2(P_{j,i}) \\
&= \begin{cases}
\frac{(a_2+a_3)\beta_{m,k} \tau /\ln 2 }{a_2 \beta_{m,k} \tau P_{m,k} + a_3 \tau \sum_{i=1}^{K_m}\beta_{m,i} P_{m,i} +a_3}, & j=m, i=k,\\
\frac{a_3\beta_{m,i} \tau /\ln 2}{a_2 \beta_{m,k} \tau P_{m,k} + a_3 \tau \sum_{i=1}^{K_m}\beta_{m,i} P_{m,i} +a_3},& j=m, i \neq k, \\
0, & j\neq m.
\end{cases}
\end{align*}

\begin{algorithm}[t]
\caption{{\small{Proposed Power Allocation Algorithm for Sum Rate Maximization}}}
\label{algorithms3}
{\bf{Initialize}}  $\varepsilon \leftarrow 10^{-3}$; Initialize feasible downlink power ${\bf{Q}}^{(0)}$; \\ 
\Repeat($\left\{\rm{Outer\;iteration}\right\}$){$\varepsilon^\star\leq \varepsilon$}
{
\textbf{Uplink power allocation}: \\
\Repeat($\left\{\rm{Inner\;iteration}\right\}$){${\bf{P}}^{(l)}$ {\rm{converges}}}
 {${\bf{P}}^{(l)} \leftarrow {\rm{max}} ~~\sum_{m=1}^M \sum_{k=1}^{K_m} [ f_1({\bf{P}})-f_2({\bf{P}}^{(l-1)})- {(P_{m,k} -P_{m,k}^{(l-1)})} \times  \sum_{j=1}^M \sum_{i=1}^{K_j} {\nabla f_2({P_{j,i}}^{(l-1)})}  ]$~${\rm{s.t.}} \; (\ref{eq:OB}\rm{b})$}
 $R_{\rm{sum}}^u \leftarrow \sum_{m=1}^M \sum_{k=1}^{K_m} R^{sec}_{m,k}$;\\
 \textbf{Downlink power allocation}: \\
\Repeat($\left\{\rm{Inner\;iteration}\right\}$){${\bf{Q}}^{(l)}$ {\rm{converges}}}
 {${\bf{Q}}^{(l)} \leftarrow {\rm{max}} ~~ \sum_{m=1}^M \sum_{k=0}^{K_m} [ g_1({\bf{Q}}) + g_3({\bf{Q}}) -g_2({\bf{Q}}^{(l-1)})- g_4({\bf{Q}}^{(l-1)}) - {(Q_{m,k} -Q_{m,k}^{(l-1)})} \times  \sum_{j=1}^M \sum_{i=0}^{K_j} {\nabla g_2({Q_{j,i}}^{(l-1)}) +\nabla g_4({Q_{j,i}}^{(l-1)})}  ] $
 
 ~~${\rm{s.t.}} \; (\ref{eq:OB}\rm{c}), (\ref{eq:OB}\rm{d})$}
$R_{\rm{sum}}^d \leftarrow \sum_{m=1}^M \sum_{k=1}^{K_m} R^{sec}_{m,k}$;\\
Compute $\varepsilon^{\star}\! \leftarrow R_{\rm{sum}}^d-R_{\rm{sum}}^u$.\\
}
{\bf{if}} $R^{sec}_{m,k} <0, \forall m\in \{1,\cdots, M\}, k\in \{1,\cdots, K_m\}$ \\
\hspace{10 pt}$R^{sec}_{m,k} \leftarrow 0$; \\
{\bf{end}} \\
$R_{\rm{sum}} \leftarrow \sum_{m=1}^M \sum_{k=1}^{K_m} R^{sec}_{m,k}$.
\end{algorithm}

The following procedure generates a sequence $\{{\bf{P}}^{(l)} \} $ of improved feasible solutions \cite{DC, N_vucic}. Initialized from a feasible $\{{\bf{P}}^{(0)} \} $, $\{{\bf{P}}^{(l)} \} $ is obtained as the optimal
solution of the following convex problem at the $l$-th iteration:
\begin{align} \label{dc_convex 1}
\underset{\bf{P}} {\rm{max}} ~~ &\sum_{m=1}^M \sum_{k=1}^{K_m} \bigg[ f_1({\bf{P}})-f_2({\bf{P}}^{(l-1)})- \n
& ~ {(P_{m,k} -P_{m,k}^{(l-1)})} \times  \sum_{j=1}^M \sum_{i=1}^{K_j} {\nabla f_2({P_{j,i}}^{(l-1)})}  \bigg] ~  \n
{\rm{s.t.}} &\; (\ref{eq:OB}\rm{b}).
\end{align}

Note that \eqref{dc_convex 1} can be efficiently solved by available convex software packages \cite{Cvx}. Moreover, since there exists no inter-cluster interference, the sum rate maximization can be done {\color{black}in} parallel for each cluster, i.e., the {\color{black}system} sum rate maximization equals to the {\color{black}cluster sum} rate maximization.

After solving the above problem, we can obtain the value for ${\bf{P}}$. Accordingly, we can obtain $\rho_{m,k}$. On this basis, for sub-problem (2), after some mathematical manipulations, $\bar{R}_{m,k} $ can be expressed as {\color{black}\eqref{SE_m_k} on the top of the next page.
\begin{figure*}[!t]
\normalsize
\begin{align}  \label{SE_m_k}
  \bar{R}_{m,k}      &=(1-\frac{\tau}{T}) \log_2\left(1+\frac{b_1 Q_{m,k}}{b_2 Q_{m,k} + b_3 \sum_{i=1}^{k-1}Q_{m,i}+ b_2 Q_{m,0} + \beta_{m,k}  \sum_{j \neq m} \sum_{i=0}^{K_j}Q_{j,i} +1}\right) \n
  &=(1-\frac{\tau}{T}) \ub{ \log_2\left({(b_1+b_2) Q_{m,k} + b_3 \sum_{i=1}^{k-1}Q_{m,i}+ b_2 Q_{m,0} + \beta_{m,k}  \sum_{j \neq m} \sum_{i=0}^{K_j}Q_{j,i} +1}\right) }_{g_1(\bf{Q})} \n
  &~ - (1-\frac{\tau}{T}) \ub{ \log_2\left({b_2 Q_{m,k} + b_3 \sum_{i=1}^{k-1}Q_{m,i}+ b_2 Q_{m,0} + \beta_{m,k}  \sum_{j \neq m} \sum_{i=0}^{K_j}Q_{j,i} +1}\right) }_{g_2(\bf{Q})}
\end{align}
\end{figure*}
Note that in \eqref{SE_m_k}, $b_1=\rho_{m,k}\beta_{m,k} N_t$, $b_2=\beta_{m,k} (1- \rho_{m,k})$, and $b_3=\beta_{m,k} (\rho_{m,k}N_t+1-\rho_{m,k})$.}

The gradient of $g_2$ at $Q_{j,i}, \forall j \in \{1, \cdots, M\}, i \in \{0, \cdots, K_j\}$ is given by {\color{black}\eqref{der_g_2} on the top of the next page. }
\begin{figure*}[!t]
\normalsize
\begin{equation} \label{der_g_2}
\nabla g_2(Q_{j,i})=
\begin{cases}
\frac{b_2 }{{b_2 Q_{m,k} + b_3 \sum_{i=1}^{k-1}Q_{m,i}+ b_2 Q_{m,0} + \beta_{m,k}  \sum_{j \neq m} \sum_{i=0}^{K_j}Q_{j,i} +1}}, & j=m, i=k ~\rm{or}~ 0,\\
\frac{b_3}{{b_2 Q_{m,k} + b_3 \sum_{i=1}^{k-1}Q_{m,i}+ b_2 Q_{m,0} + \beta_{m,k}  \sum_{j \neq m} \sum_{i=0}^{K_j}Q_{j,i} +1}},& j=m,  i=1, \cdots, k-1, \\
\beta_{m,k}, & j\neq m, \\
0, & {\color{black}\rm{otherwise}}.
\end{cases}
\end{equation}
\end{figure*}

Next, let us consider $-\bar{R}^{e}_{m,k}$, which can be re-written as
\begin{align}
  &-\bar{R}_{m,k}^{e}=(1-\frac{\tau}{T}) \ub{ \log_2\left({\beta_E \sum_{i \neq k} Q_{m,i}+ \beta_E  \sum_{j \neq m} \sum_{i=0}^{K_j}Q_{j,i} +1}\right) }_{g_3(\bf{Q})} \n
  &~~~ - (1-\frac{\tau}{T}) \ub{ \log_2\left({\beta_E \sum_{i=0}^{K_m} Q_{m,i}+ \beta_E  \sum_{j \neq m} \sum_{i=0}^{K_j}Q_{j,i} +1}\right) }_{g_4(\bf{Q})}.
\end{align}

The gradient of $g_4$ at $Q_{j,i}, \forall j \in \{1, \cdots, M\}, i \in \{0, \cdots, K_j\}$ is given by
\begin{equation}
\nabla g_4(Q_{j,i})= \frac{\beta_E /\ln 2}{ \beta_E \sum_{i=0}^{K_m} Q_{m,i}+ \beta_E  \sum_{j \neq m} \sum_{i=0}^{K_j}Q_{j,i} +1 }.
\end{equation}

The following procedure generates a sequence $\{{\bf{Q}}^{(l)} \} $ of improved feasible solutions {\color{black}\cite{DC, N_vucic}}. Initialized from a feasible $\{{\bf{Q}}^{(0)} \} $, $\{{\bf{Q}}^{(l)} \} $ is obtained as the optimal
solution of the following convex problem at the $l$-th iteration:
\begin{align} \label{dc_convex_2}
&\underset{\bf{Q}} {\rm{max}}  \sum_{m=1}^M \sum_{k=0}^{K_m} [ g_1({\bf{Q}}) + g_3({\bf{Q}}) -g_2({\bf{Q}}^{(l-1)})- g_4({\bf{Q}}^{(l-1)}) - \n
&{(Q_{m,k} -Q_{m,k}^{(l-1)})} \times   \sum_{j=1}^M \sum_{i=0}^{K_j} {\nabla g_2({Q_{j,i}}^{(l-1)}) +\nabla g_4({Q_{j,i}}^{(l-1)})}  ] ~  \nonumber \\
&{\rm{s.t.}} \; (\ref{eq:OB}\rm{c}), (\ref{eq:OB}\rm{d}) .  
\end{align}

Note that \eqref{dc_convex_2} can also be efficiently solved by available convex software packages \cite{Cvx}.

Now we have solved the two sub-problems. We repeat them after each other until convergence. {\color{black}Then, for those users with negative rates, we set their rates to zero {\color{black}following the $[\cdot]^+$ operation}.} {\color{black}The specific procedure is summarized in \text{Algorithm 1}.}

\subsection{EE Maximization}
It is clear that \eqref{eq:EE} belongs to a fractional problem, which can be transformed into a series of parametric subtractive-form subproblems as {\color{black}\eqref{eq:EE_SE} on the top of the next page {\color{black}based on Dinkelbach algorithm \cite{W_dink}}. 
\begin{figure*}[!t]
\normalsize
\begin{equation}\label{eq:EE_SE}
\displaystyle {\underset{\bf{P}, \bf{Q}}{\rm{max}} }~ {\sum_{m=1}^M \sum_{k=1}^{K_m} R^{sec}_{m,k}} - \lambda^{(l-1)} \left({\sum_{m=1}^M \sum_{k=1}^{K_m} P_{m,k}+\sum_{m=1}^M \sum_{k=0}^{K_m} Q_{m,k} +P_f} \right),   ~{\rm{s.t.}}~(\ref{eq:OB}\rm{b})-(\ref{eq:OB}\rm{d}).
\end{equation}
\end{figure*}

Note in \eqref{eq:EE_SE}, $\lambda^{(l-1)}$ is a non-negative parameter. Starting from $\lambda^{(0)}=0$, $\lambda^{(l)}$ can be updated by $\lambda^{(l)} = \frac{{\sum_{m=1}^M \sum_{k=1}^{K_m} R^{sec}_{m,k}}^{(l)}}{\sum_{m=1}^M \sum_{k=1}^{K_m} P_{m,k}^{(l)}+\sum_{m=1}^M \sum_{k=0}^{K_m} Q_{m,k}^{(l)} +P_f}$, where ${R^{sec}_{m,k}}^{(l)}$,  $P_{m,k}^{(l)} $ and $Q_{m,k}^{(l)}$ are the updated rates and power after solving \eqref{eq:EE_SE}. {\color{black}As shown in \cite{W_dink}, $\lambda^{(l)}$ keeps growing as $l$ increases. When $\lambda^{(l)}-\lambda^{(l-1)} $ is smaller than a certain threshold, e.g., $ 10^{-3}$, the iterations terminate, and the obtained $\lambda^{(l)}$ is the maximum EE of \eqref{eq:EE}}.}



Then, the problem lies in how to solve \eqref{eq:EE_SE} for a given $\lambda$. It is clear that \eqref{eq:EE_SE} is similar to the sum rate maximization problem \eqref{eq:OB}, except for the extra linear part in the objective function. Adding a linear part does not affect the way of solving the problem, and thus, we can apply the proposed sum rate maximization here directly. The specific procedure is summarized in \text{Algorithm 2}.

\begin{algorithm}[t]
\caption{{\small{Energy-Efficient Power Allocation Algorithm}}}
\label{algorithms3}
{\bf{Initialize}}  $\varepsilon \leftarrow 10^{-6}$, $\lambda \leftarrow 0$, Initialize feasible power ${\bf{P}}^{(0)}$. \\
\Repeat($\left\{\rm{Outer\;iteration}\right\}$){$\varepsilon^\star\leq \varepsilon$}
{
\Repeat($\left\{\rm{Inner\;iteration}\right\}$){${\bf{P}}^{(l)}, {\bf{Q}}^{(l)}$ {\rm{converge}}}
 {${\bf{P}}^{(l)}, {\bf{Q}}^{(l)} \leftarrow {\rm{max}} ~~{\sum_{m=1}^M \sum_{k=1}^{K_m} R^{sec}_{m,k}} - \lambda ({\sum_{m=1}^M \sum_{k=1}^{K_m} P_{m,k}+\sum_{m=1}^M \sum_{k=0}^{K_m} Q_{m,k} +P_f})$ 
 
 ~~${\rm{s.t.}} \; (\ref{eq:OB}\rm{b}), (\ref{eq:OB}\rm{c}),(\ref{eq:OB}\rm{d})$}
Compute $\varepsilon^{\star}\! \leftarrow \!{\sum_{m=1}^M \sum_{k=1}^{K_m} R^{sec}_{m,k}}^{(l)} - \lambda ({\sum_{m=1}^M \sum_{k=1}^{K_m} P_{m,k}^{(l)}+\sum_{m=1}^M \sum_{k=0}^{K_m} Q_{m,k}^{(l)} +P_f})$.\\
Update $\lambda  \leftarrow \frac{{\sum_{m=1}^M \sum_{k=1}^{K_m} R^{sec}_{m,k}}^{(l)}}{\sum_{m=1}^M \sum_{k=1}^{K_m} P_{m,k}^{(l)}+\sum_{m=1}^M \sum_{k=0}^{K_m} Q_{m,k}^{(l)} +P_f}$.\\
}
{\bf{if}} ${R^{sec}_{m,k}}^{(l)} <0, \forall m\in \{1,\cdots, M\}, k\in \{1,\cdots, K_m\}$ \\
\hspace{10 pt}${R^{sec}_{m,k}}^{(l)} \leftarrow 0$; \\
{\bf{end}} \\
$\eta_{\rm{EE}}  \leftarrow \frac{{\sum_{m=1}^M \sum_{k=1}^{K_m} R^{sec}_{m,k}}^{(l)}}{\sum_{m=1}^M \sum_{k=1}^{K_m} P_{m,k}^{(l)}+\sum_{m=1}^M \sum_{k=0}^{K_m} Q_{m,k}^{(l)} +P_f}$.\\
\end{algorithm}

{\color{black}
\subsection{Complexity and Convergence}
The proposed SE maximization algorithm includes inner and outer iterations. For the inner iteration, i.e., the DC programming,
its convergence has been shown in \cite{N_vucic, DC}. For the outer iteration, on one hand, the SE increases or remains unchanged for both the uplink and downlink {\color{black}PA}; on the other hand, there exists an upper bound for the SE. Therefore, the outer iteration terminates within a limited number of iterations, i.e., the proposed SE maximization algorithm always converges.

The proposed EE maximization algorithm also includes inner and outer iterations. For the inner iteration, i.e., the SE maximization,
its convergence has been shown above. For the outer iteration, i.e., the fractional programming, it always converges to the stationary and optimal solution~\cite{W_dink}. Therefore, the proposed EE maximization algorithm always converges.

Now, we discuss {\color{black}the computational complexity of the proposed algorithms}. First, we look at the proposed SE maximization algorithm. Denote the number of iterations for solving the uplink and downlink {\color{black}PA} as $I_1$ and $I_2$, respectively. The corresponding number of dual variables for solving \eqref{dc_convex 1} and \eqref{dc_convex_2} is denoted as $D_1$ and $D_2$, respectively. Then, if the number of outer iteration is $I_3$, the overall computational complexity of the proposed SE maximization algorithm is $O \left(I_3 (I_1D_1^2+ I_2 D_2^2) \right)$. Next, we consider the proposed EE maximization problem. Denote its outer iteration as $I_4$, then {\color{black}it can be easily shown that} the overall computational complexity of the proposed EE maximization algorithm is $O \left(I_4 I_3 (I_1D_1^2+ I_2 D_2^2) \right)$.
}

\begin{figure}[h]
\centering
\includegraphics[width=0.65\textwidth]{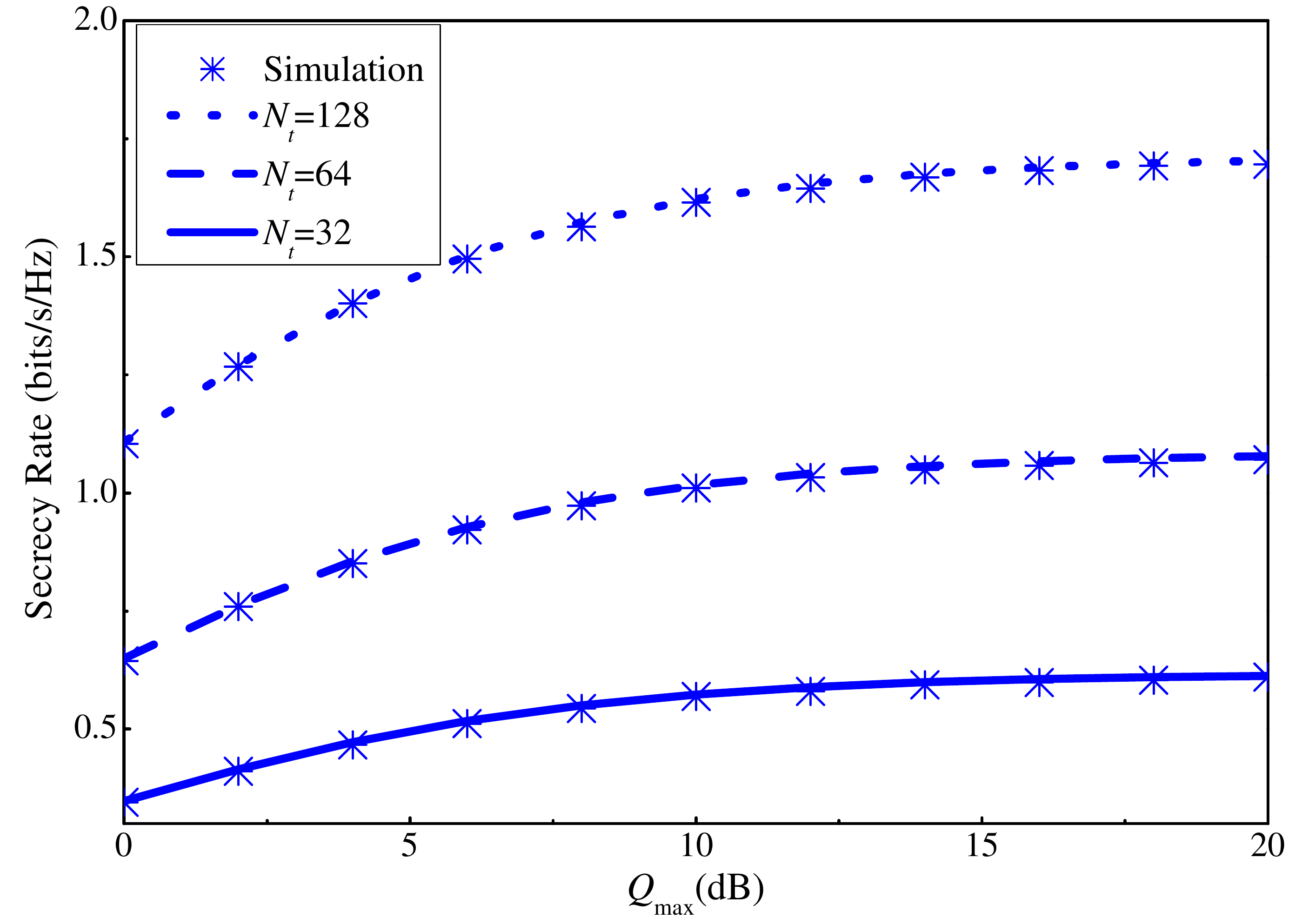}
\caption{{Secrecy rate at the $2^{nd}$ user in the $5^{th}$ cluster versus the total transmit power at the BS, for different numbers of transmit antennas.}}
\label{QvsRs}
\end{figure}

\section{Numerical Results}
In this section, we firstly investigate the behavior of the system without PA to highlight the effects of key parameters on the secrecy performance in subsection~\ref{Fixed PA}.
The effectiveness of our proposed PA algorithms is {\color{black}then} evaluated in subsection~\ref{Optimized PA}.
\subsection{Fixed PA}\label{Fixed PA}
\begin{figure}
\centering
\includegraphics[width=0.65\textwidth]{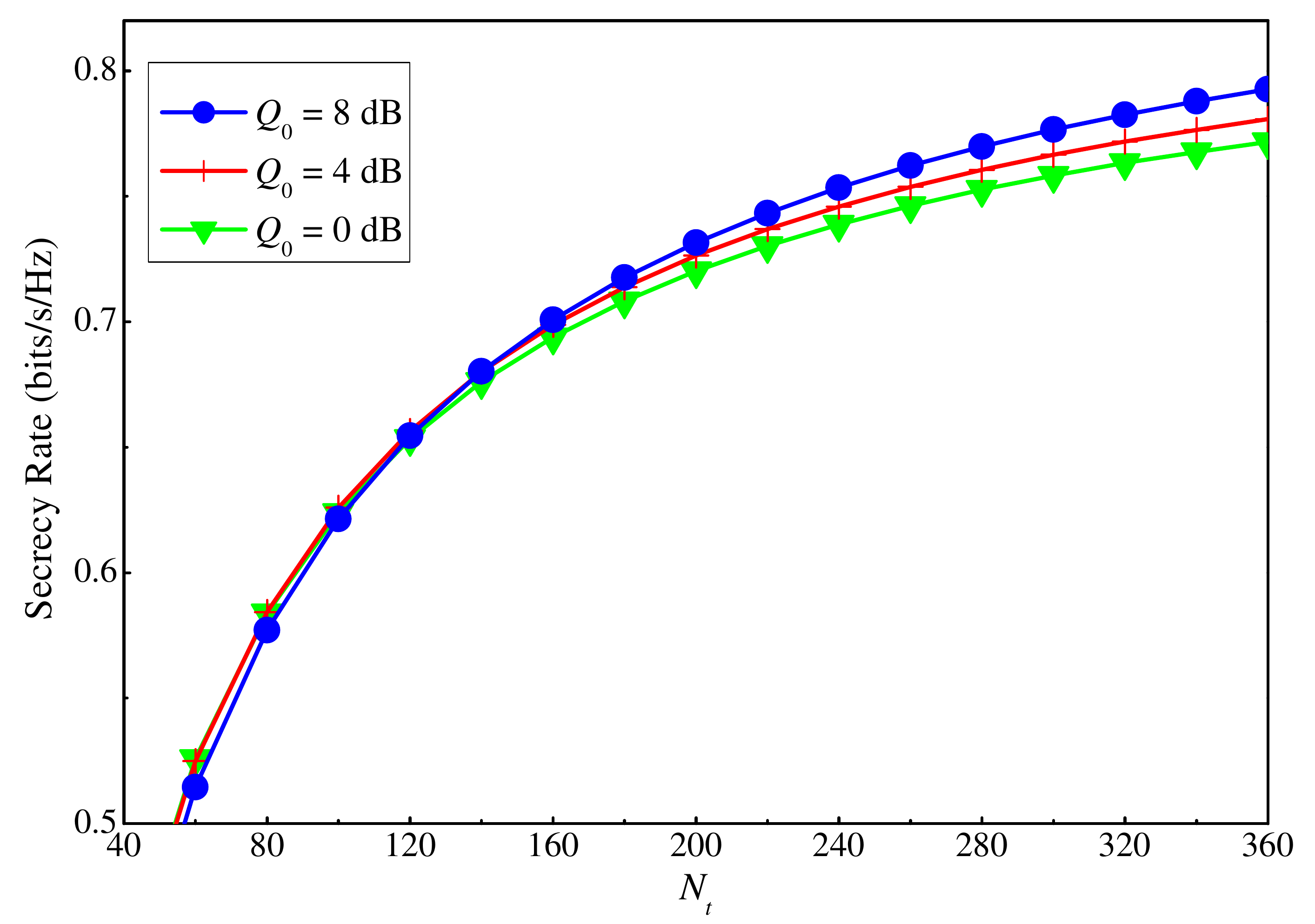}
\caption{{Secrecy rate at the $2^{nd}$ user in the $5^{th}$ cluster versus the number of antennas at the BS, for different AN powers.}}
\label{NtvsRsVarAN}
\end{figure}

Without loss of generality, we consider the following scenario.
The total transmit power is allocated $80\%$ for information transmission and $20\%$ for AN.
The effect of varying {\color{black}the} AN power allocation will be shown later in Fig.~\ref{NtvsRsVarAN}.
The power is equally assigned to each user, and the AN power for each cluster is the same.
$T=300$ units and $\tau = M$ units. $\beta_{m,k}$ for each user is a random value between 0 and 100 and satisfies the condition $\beta_{m,1}\geq ... \geq \beta_{m,K_m}$, while that for the illegitimate user is fixed to $\beta_{E}=$10.
Unless explicitly mentioned, this setup is kept throughout the section.

Without loss of generality, the ergodic secrecy rate of the $2^{nd}$ user in the $5^{th}$ cluster is selected to show in Fig.~\ref{QvsRs}.
The number of cluster is $M=10$ and the number of users in a cluster is $K=2$.\footnote{The subscript $m$ in $K_m$ is dropped since the same number of users is considered in clusters.}
It can be seen that the approximation in \eqref{secrate_approx} and the simulation results match very well.
Throughout the numerical results section, this approximation will be used.
Besides, when the total transmit power at the BS increases, the secrecy rate at a user converges to a constant value.
This is because of the interference and AN within the cluster and from other clusters.
In addition, we can also observe that an increase in the number of antennas at the BS can lift the secrecy performance.
The reason is that by increasing the number of antennas, the spatial transmitting beams become sharper, which leads to a decrease in inter-cluster interference and AN leakage, and an increase in the desired signal.
The next figure will reveal how to take advantage of this property to enhance secrecy performance.

\begin{figure}
\centering
\includegraphics[width=0.65\textwidth]{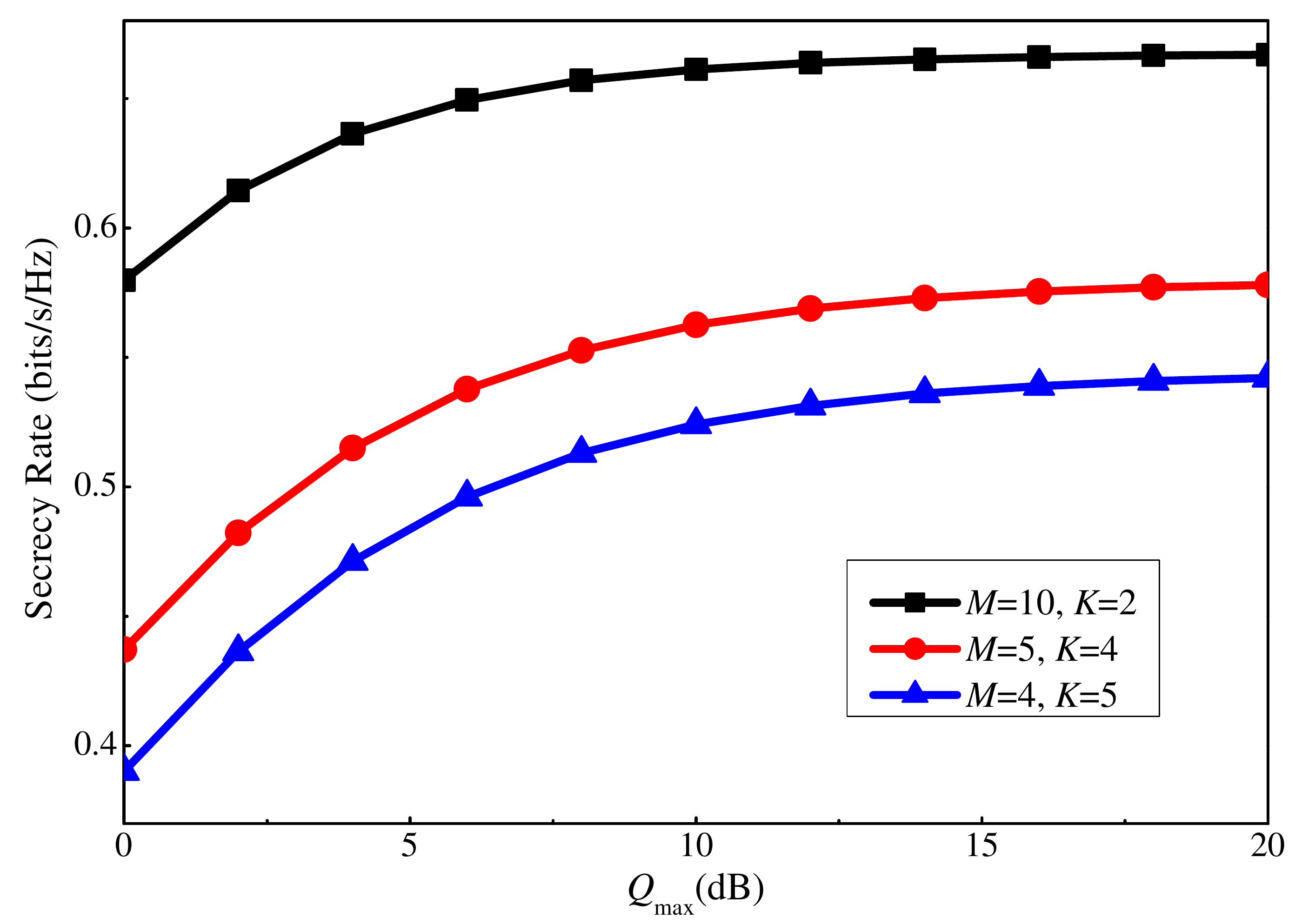}
\caption{{Secrecy rate at the $2^{nd}$ user of the $2^{nd}$ cluster versus the total transmit power at the BS, for different clustering scenarios.}}
\label{DiffKM}
\end{figure}

\begin{figure}
\centering
\includegraphics[width=0.65\textwidth]{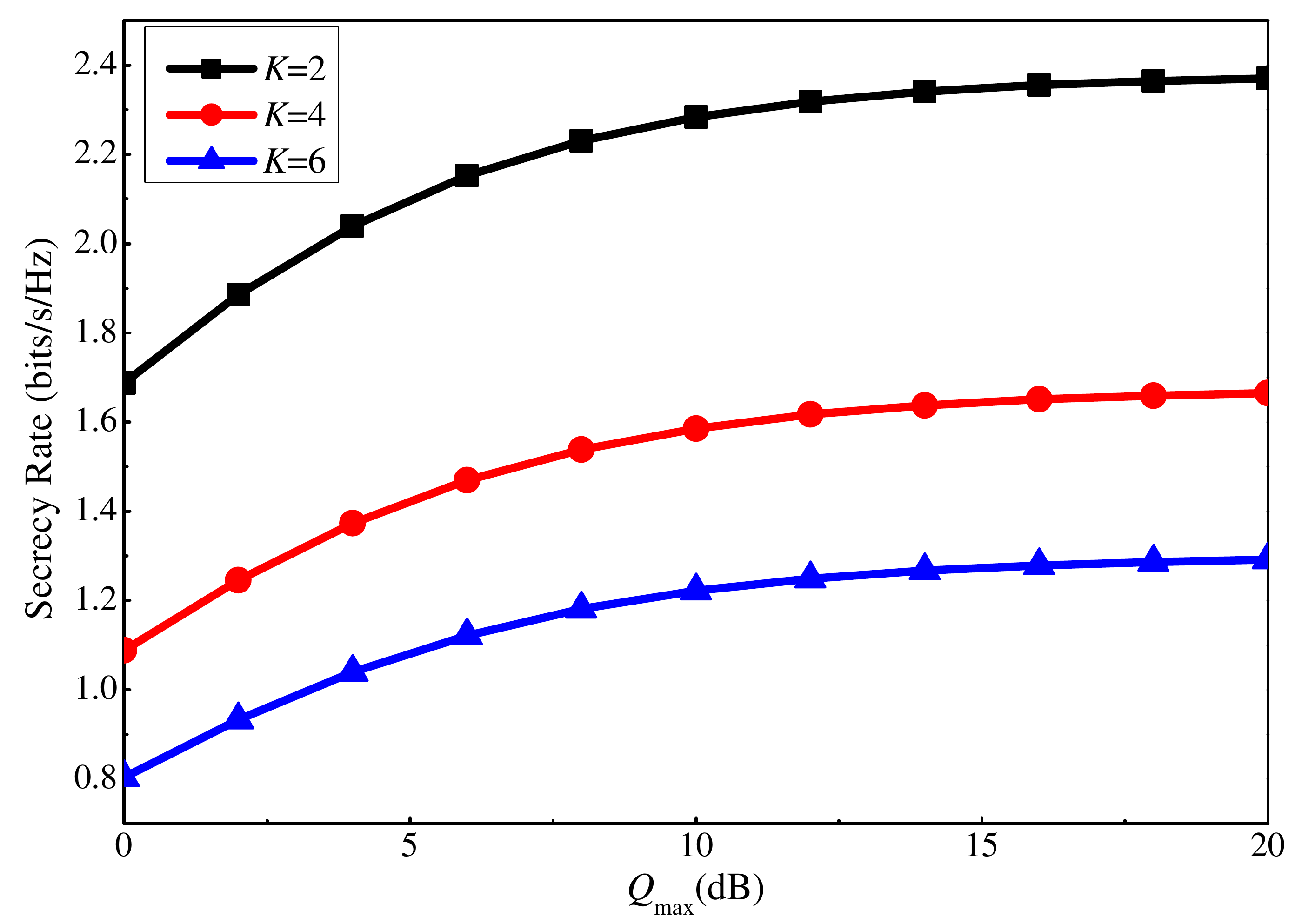}
\caption{{The secrecy rate at the the $2^{nd}$ cluster versus the total transmit power at the BS, for a fixed number of clusters and different numbers of users.}}
\label{DiffK_sumrate}
\end{figure}

In Fig.~\ref{NtvsRsVarAN}, we demonstrate the advantage of combining AN and massive MIMO technique in NOMA networks.
In this setup, the transmit power assigned to each user is $10$ dB, and the AN power is varied as $\{0, 4, 8\}$ dB.
The number of clusters is $M=10$ with $K=2$ users in a cluster.
We can observe that when the number of transmit antennas is sufficiently large, the more the AN allocated power is, the better the secrecy performance at the $2^{nd}$ user is.
The main reason is that for the legitimate side, the channel hardening property of massive MIMO technique helps reducing the AN leakage and the inter-cluster interference at each cluster.
Meanwhile, the secrecy performance of the eavesdropper decreases when the AN power increases.

{\color{black}Figures~\ref{DiffKM} and \ref{DiffK_sumrate}} depict the effect of clustering on the secrecy performance.
In Fig.~\ref{DiffKM}, the total number of users is $20$, which are clustered into three scenarios: $\{M=10, K=2\}$, $\{M=5, K=4\}$, and $\{M=4, K=5\}$.
The total transmit power for each scenario is the same.
The results show that the smaller the number of users in a cluster is, the better the secrecy performance at a user is.
Meanwhile, in Fig.~\ref{DiffK_sumrate}, the scenario of limited number of orthogonal sequences is shown.
In this scenario, we assume that the number of available orthogonal sequences is $10$, therefore, the number of clusters is $M=10$.
The number of users in a cluster is varied as $K=\{2,4,6\}$ to highlight its effect on the secrecy performance of a cluster.
It is observed that although the transmit power for each user is identical and the AN power for each cluster is the same, the cluster with more users has smaller total secrecy rate than the ones with a smaller number of users.
The reason is that when the number of users in a cluster is small, the error of the uplink training process at this cluster is also small.
As a consequence, the beam of the BS for this cluster is more precise, followed by a decrease in intra-cluster interference and AN leakage.
This also reduces the imposed interference from this cluster to other clusters.
In other words, for a better secrecy performance of each user and cluster, it is crucial to keep the number of users in a cluster small (minimum is two users for NOMA networks).

\begin{figure}
\centering
\includegraphics[width=0.65\textwidth]{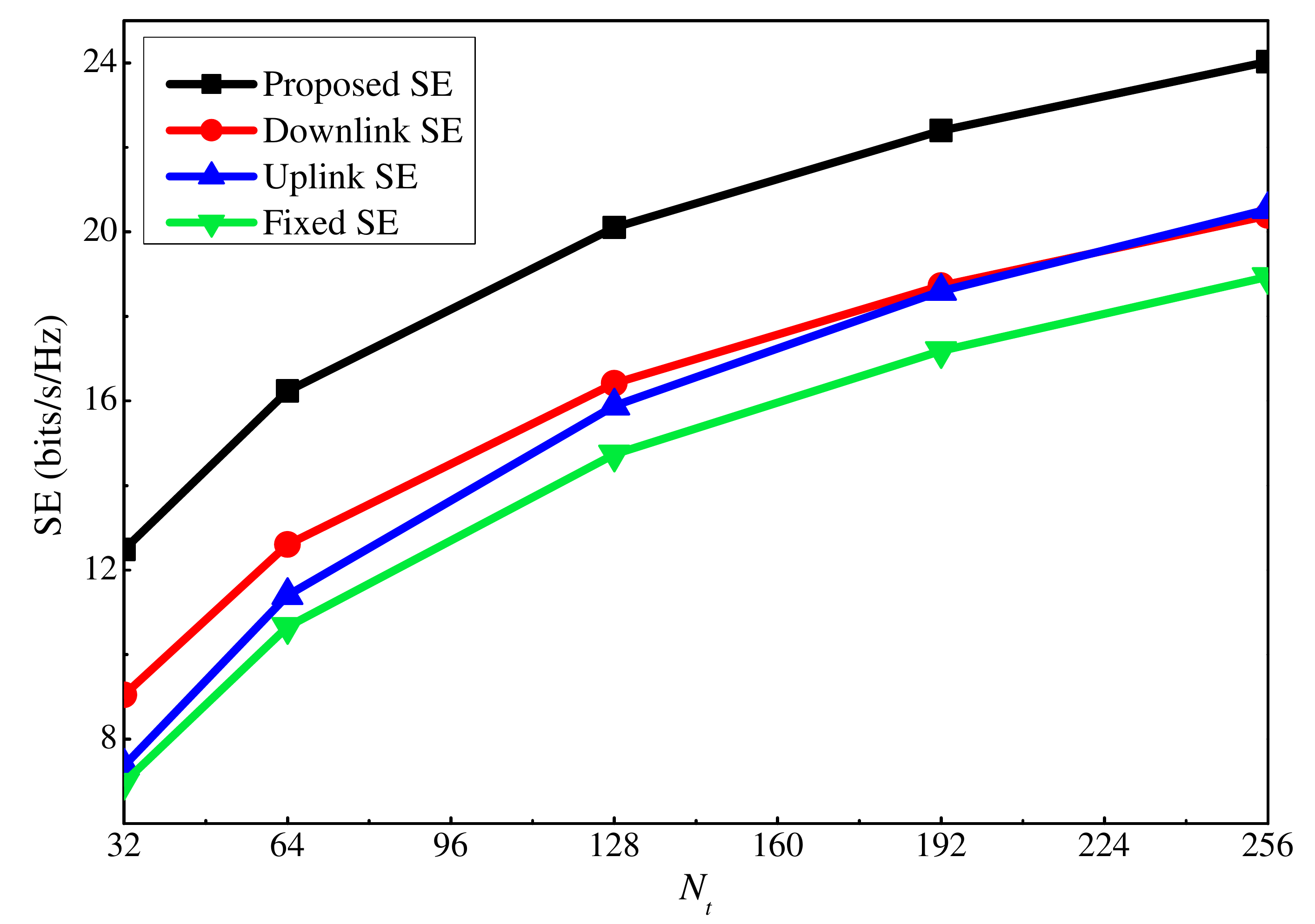}
\caption{{SE comparison between the proposed algorithm and baseline algorithms.}}
\label{SEvsNt}
\end{figure}

\begin{figure}
\centering
\includegraphics[width=0.65\textwidth]{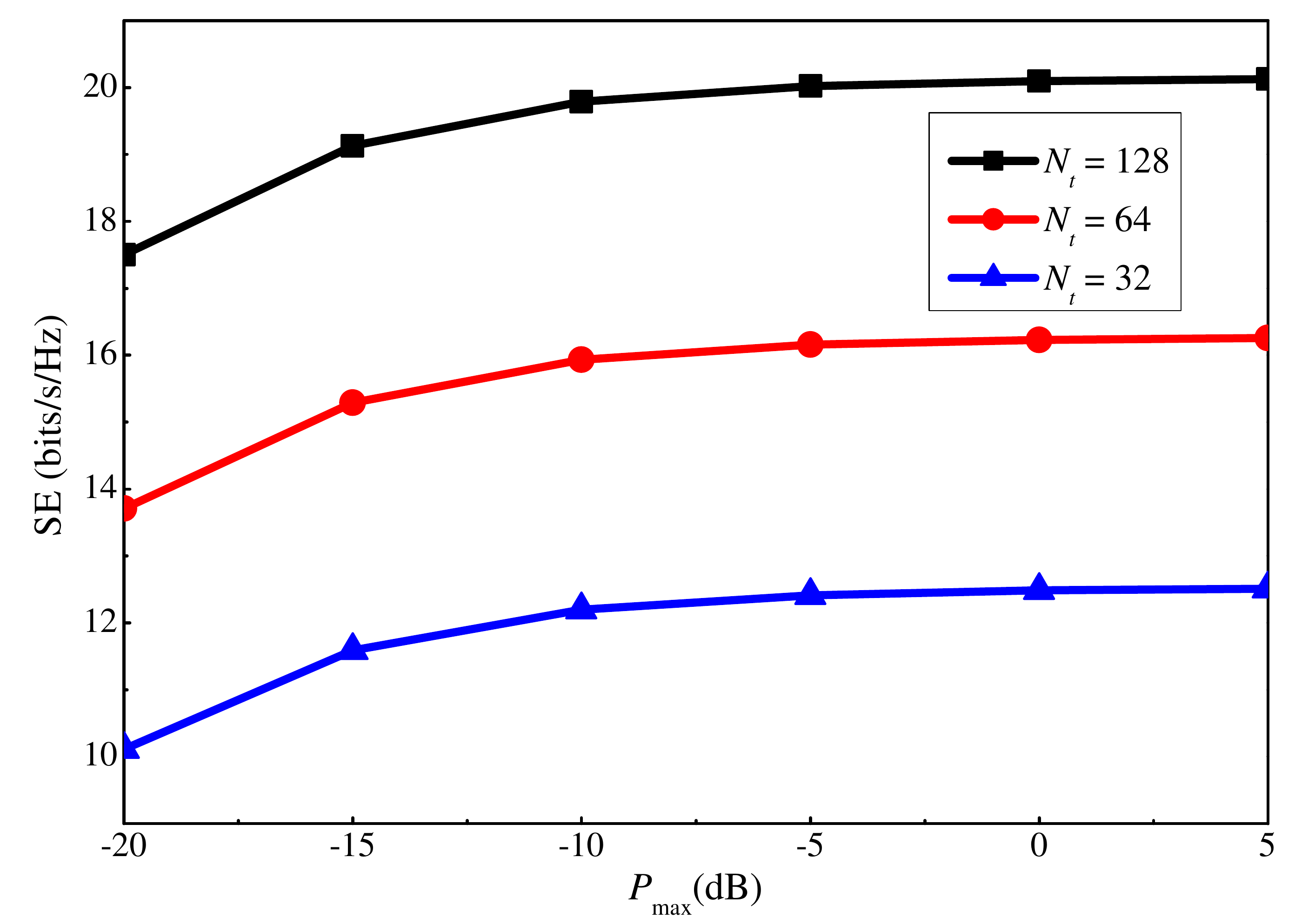}
\caption{{SE versus the maximum uplink power, for different numbers of BS antennas.}}
\label{SEvsPUL}
\end{figure}

\subsection{Optimized PA}\label{Optimized PA}
In the following, we investigate the effectiveness of the proposed SE and EE maximization algorithms. We consider a scenario with four clusters, each with three users, i.e., $M=4$ and $K=3$. The simulation parameters are as follows: $Q_{\rm{max}}=20$ dB, $P_{\rm{max}}=0$ dB. $T=300$ units. The large scale channel gain $\beta_{m,k}$ for each user is a random value between 0 and 100, while that for the illegitimate user is fixed to $\beta_{E}=$10.

First, we investigate the effectiveness of the proposed SE maximization algorithm, referred to as Proposed SE.
We compare it with three baseline algorithms, as follows: Downlink SE, which allocates the maximum uplink power to each user, and on this basis, performs PA for the downlink transmission as the Proposed SE.
In contrast, the Uplink SE first allocates $80\%$ of the total downlink power to the users equally, and $20\%$ of the total downlink power to the AN equally. Then, uplink power is optimized as the Proposed SE.
Fixed SE allocates the maximum uplink power for each user, equal downlink power allocation among the users, and the AN as in the above subsection.
As shown in Fig.~\ref{SEvsNt}, the SE provided by all four algorithms grows with the number of transmit antennas.
Moreover, among them, it can be seen that Proposed SE achieves the best performance, followed by Downlink SE, Uplink SE, and Fixed PA.
This fully reveals the necessity of performing power optimization for the considered system. Furthermore, both uplink and downlink PA are required to achieve the best performance. Nonetheless, by comparing Downlink SE and Uplink SE, we can conclude that an appropriate allocation of the downlink power may play a larger role in the current setting.

\begin{figure}
\centering
\includegraphics[width=0.65\textwidth]{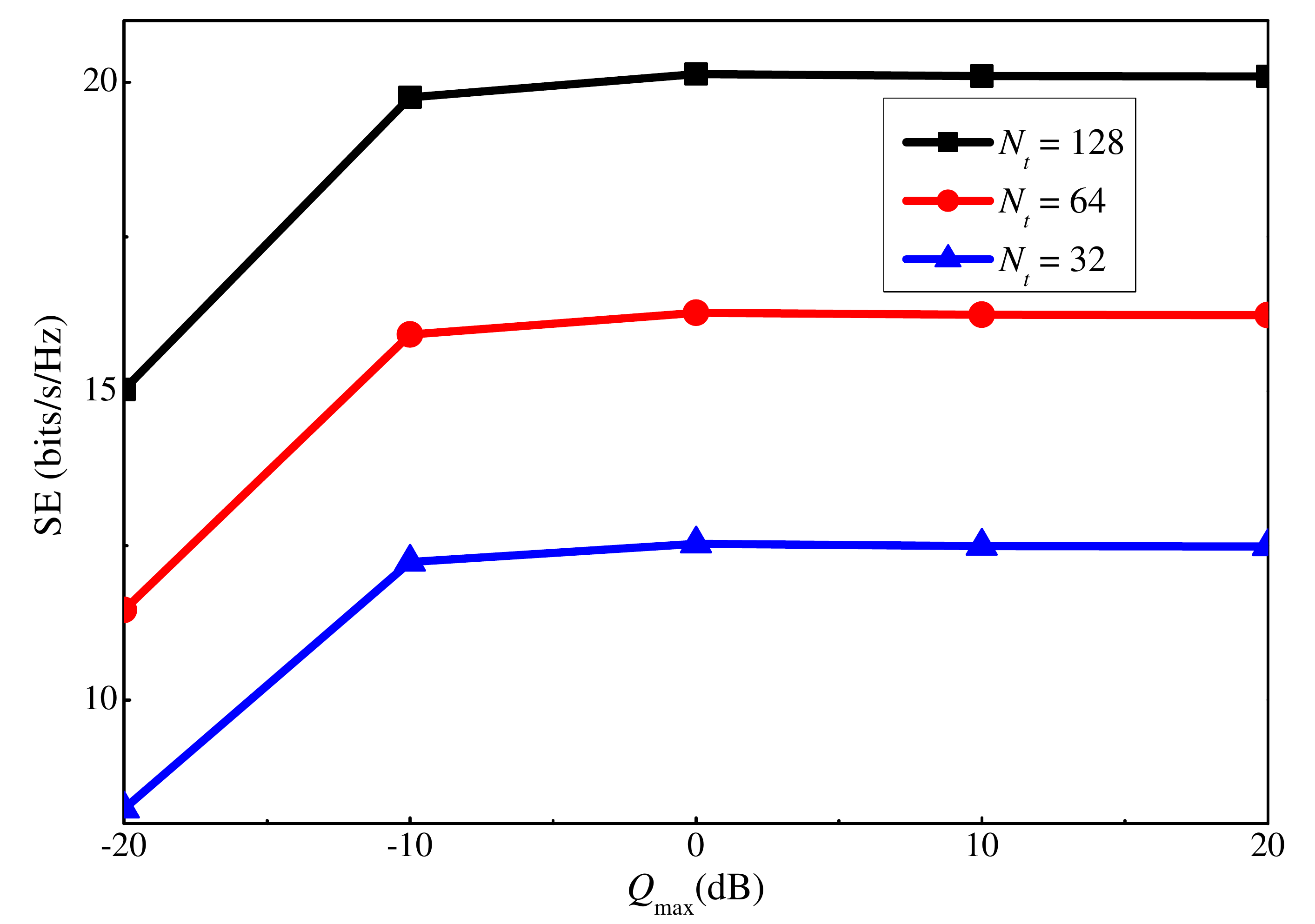}
\caption{{SE versus the maximum downlink power, for different numbers of BS antennas.}}
\label{SEvsPDL}
\end{figure}

To further show the effect of the uplink and downlink power on the achieved SE, Figs.~\ref{SEvsPUL} and \ref{SEvsPDL} plot the SE versus the maximum uplink and downlink power, respectively. $N_t=\{32, 64, 128\}$ is respectively considered in each case.
It is clear that the SE increases with both the maximum uplink and downlink powers.
The former is because increasing the maximum uplink power leads to a more precise channel estimation result, which improves the beamforming sharpness and thus, the SE.
The latter is because more power is available for data transmission.
However, after a certain point, the increase becomes minor for both power values.
This can be explained by the logarithmic relation between the power and user rate. Moreover, for {\color{black}the} downlink power, increasing it also leads to a larger illegitimate rate and intra-cluster interference.
Besides, by comparing the three antenna scenarios, we can conclude that increasing the number of antennas can significantly increase the SE.
\begin{figure}
\centering
\includegraphics[width=0.65\textwidth]{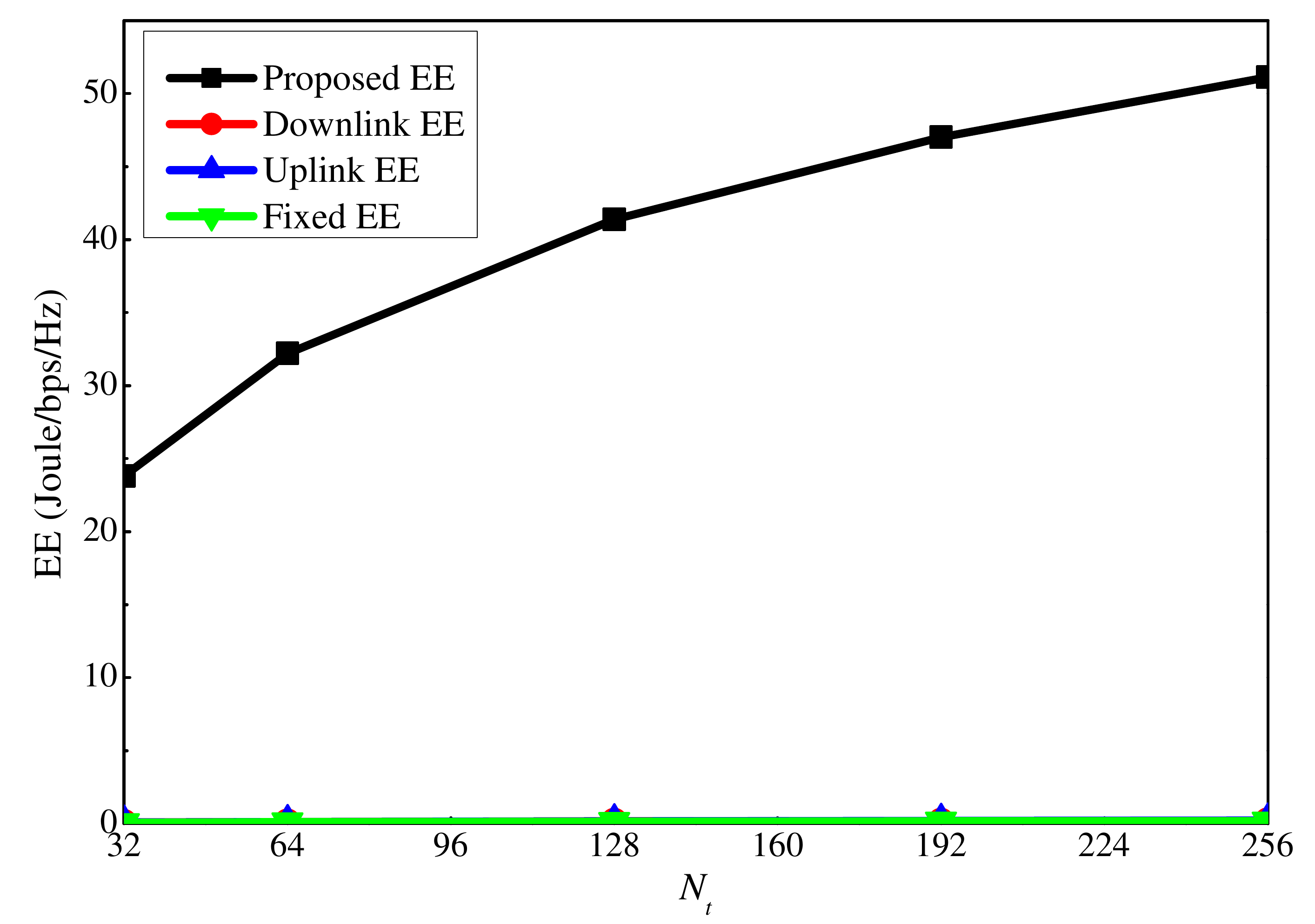}
\caption{{EE comparison between the proposed algorithm and other baseline algorithms.}}
\label{EEvsNt1}
\end{figure}

\begin{figure}
\centering
\includegraphics[width=0.65\textwidth]{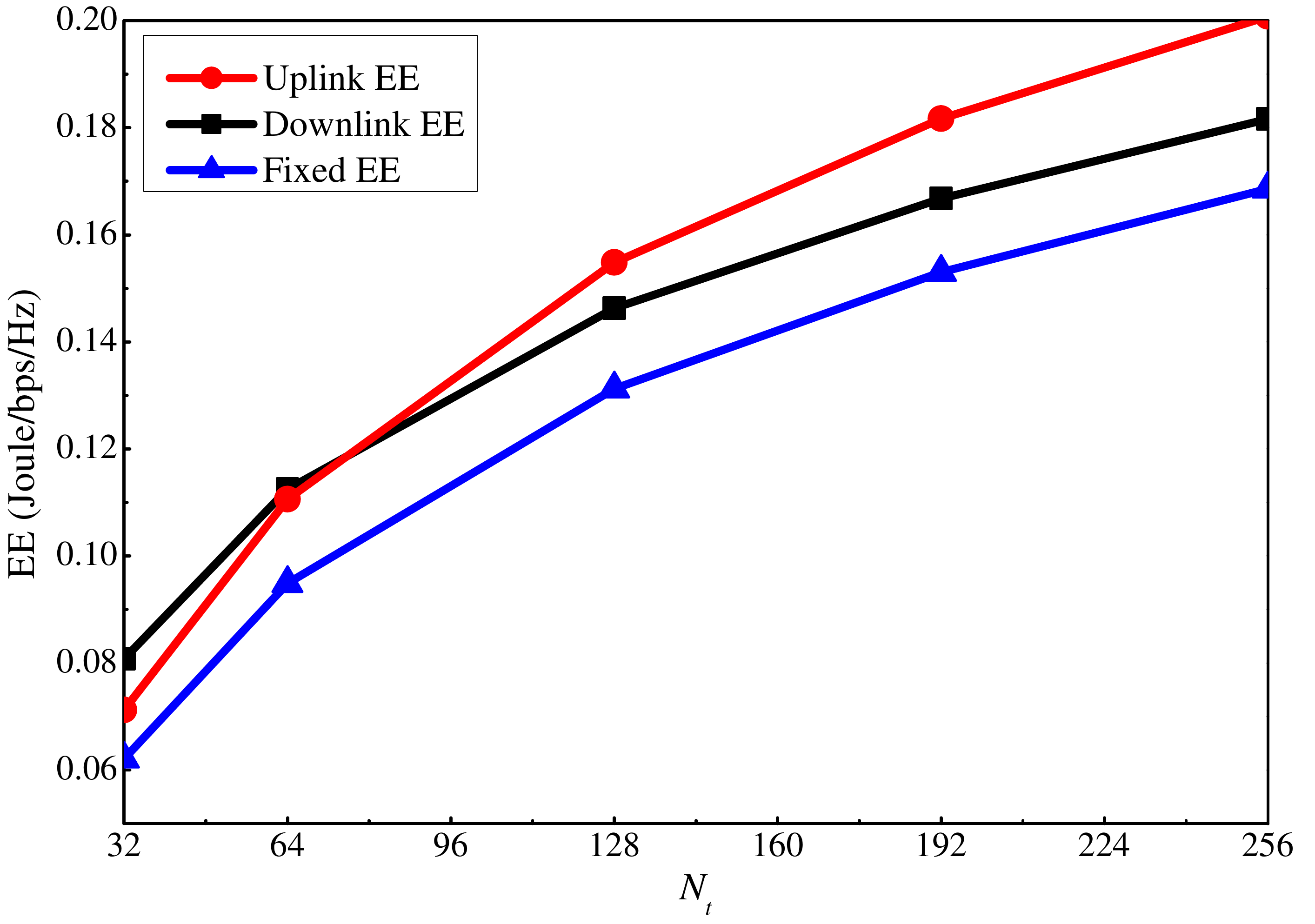}
\caption{{EE for the three baseline algorithms.}}
\label{EEvsNt2}
\end{figure}


Next, we investigate the proposed EE algorithm. Here $P_f= -5$ dB. We first compare the proposed EE maximization algorithm with the other three baseline algorithms when $Q_{\rm{max}}=20$ dB and $P_{\rm{max}}=0$ dB. According to Fig.~\ref{EEvsNt1}, the EE for the other algorithms is quite small compared with the proposed algorithm. This is because when $Q_{\rm{max}}=20$ dB and $P_{\rm{max}}=0$ dB, the power level is quite high, and thus, a large part of the available power is not used to maximize the EE. However, for the three baseline algorithms, at least one of the uplink and downlink power is fully consumed according to the setting. This leads to low EE. Figure \ref{EEvsNt2} only shows these three algorithms, and it can be seen that all of them increase with the antenna number as the proposed algorithm.

\begin{figure}
\centering
\includegraphics[width=0.65\textwidth]{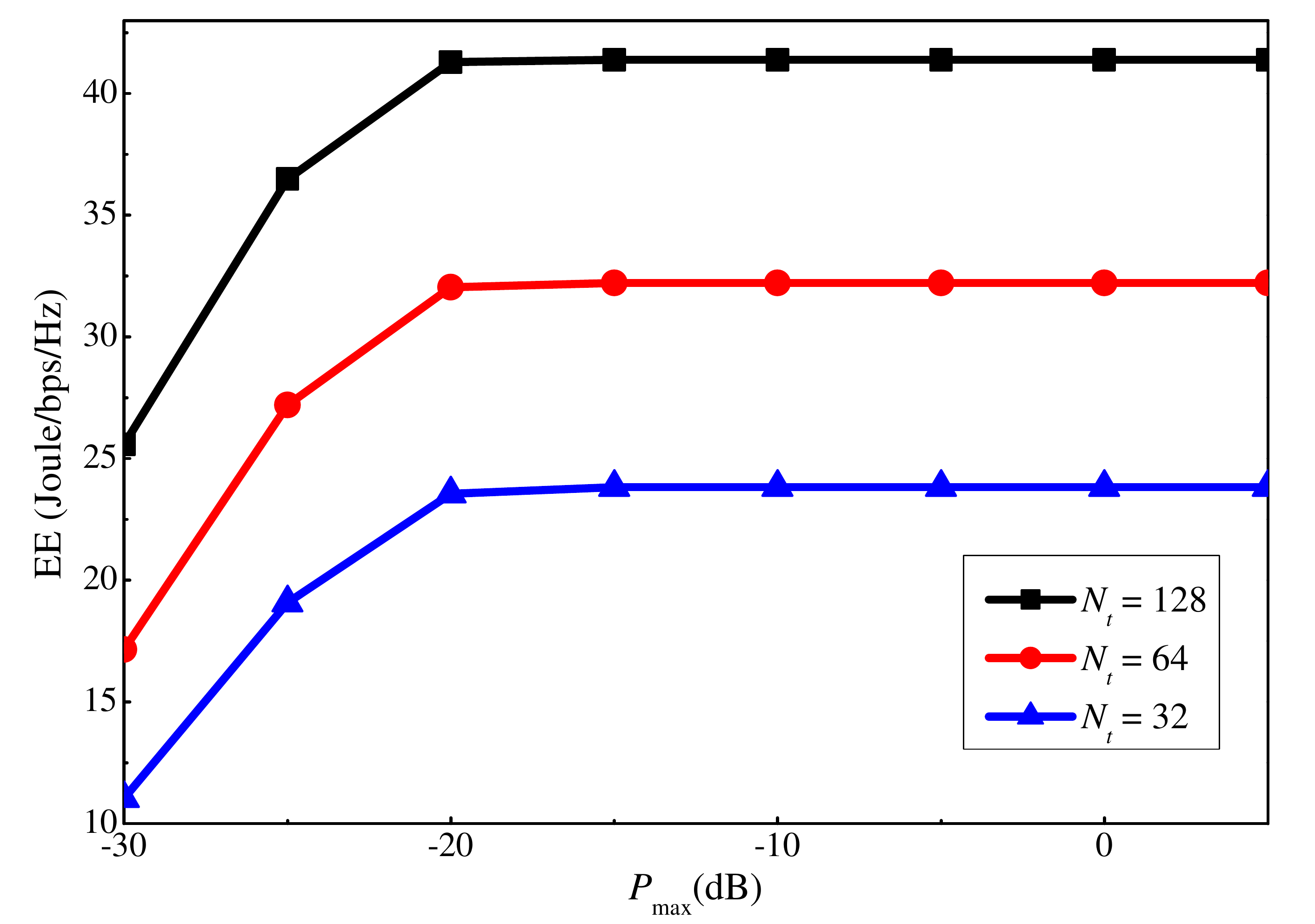}
\caption{{EE versus the maximum uplink power, for different numbers of BS antennas.}}
\label{EEvsPmax}
\end{figure}

\begin{figure}
\centering
\includegraphics[width=0.65\textwidth]{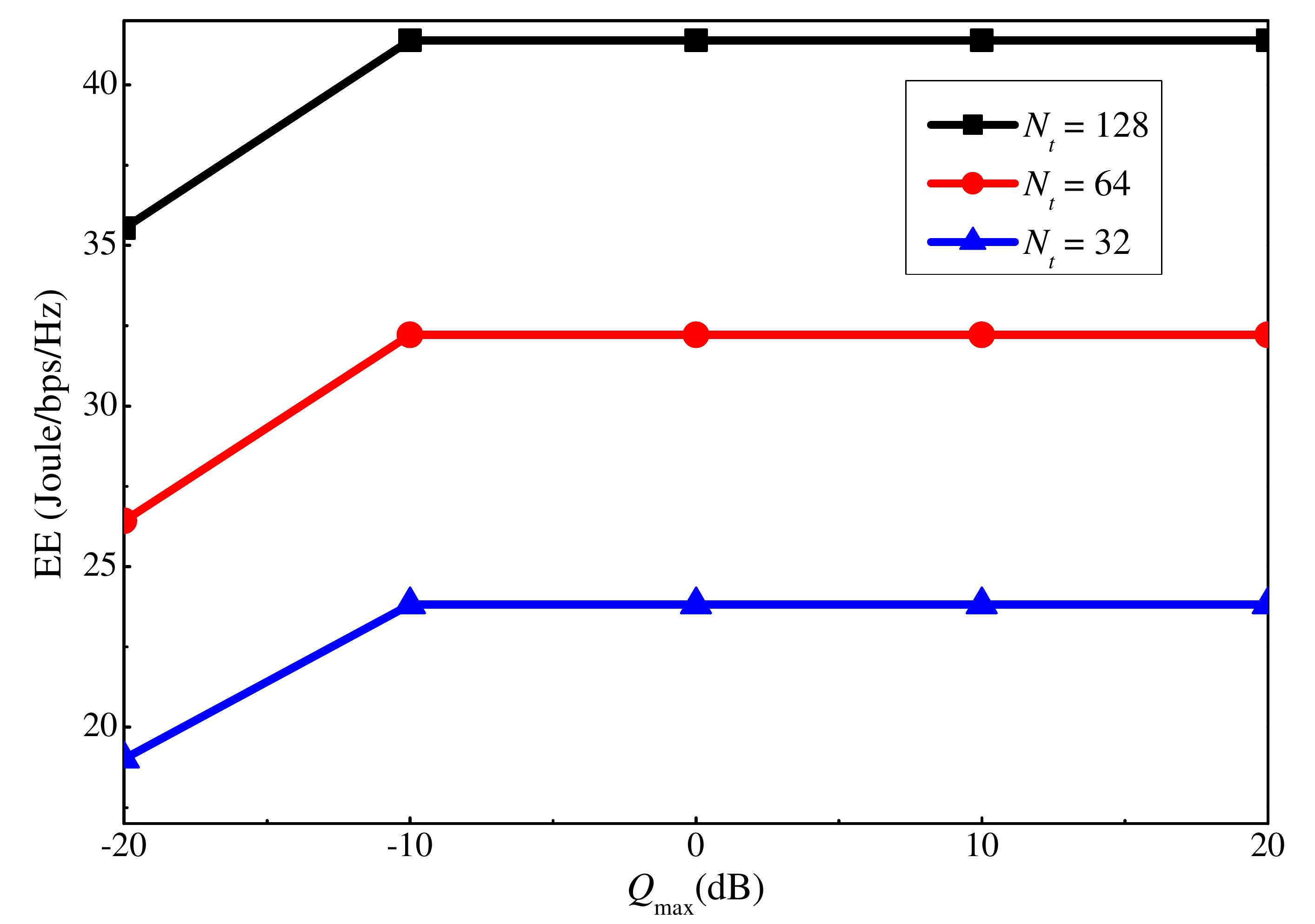}
\caption{{EE versus the maximum downlink power, for different numbers of BS antennas.}}
\label{EEvsQmax}
\end{figure}

\begin{figure*}
\centering
\begin{subfigure}{0.5\textwidth}
  \centering
  \includegraphics[width=1\linewidth]{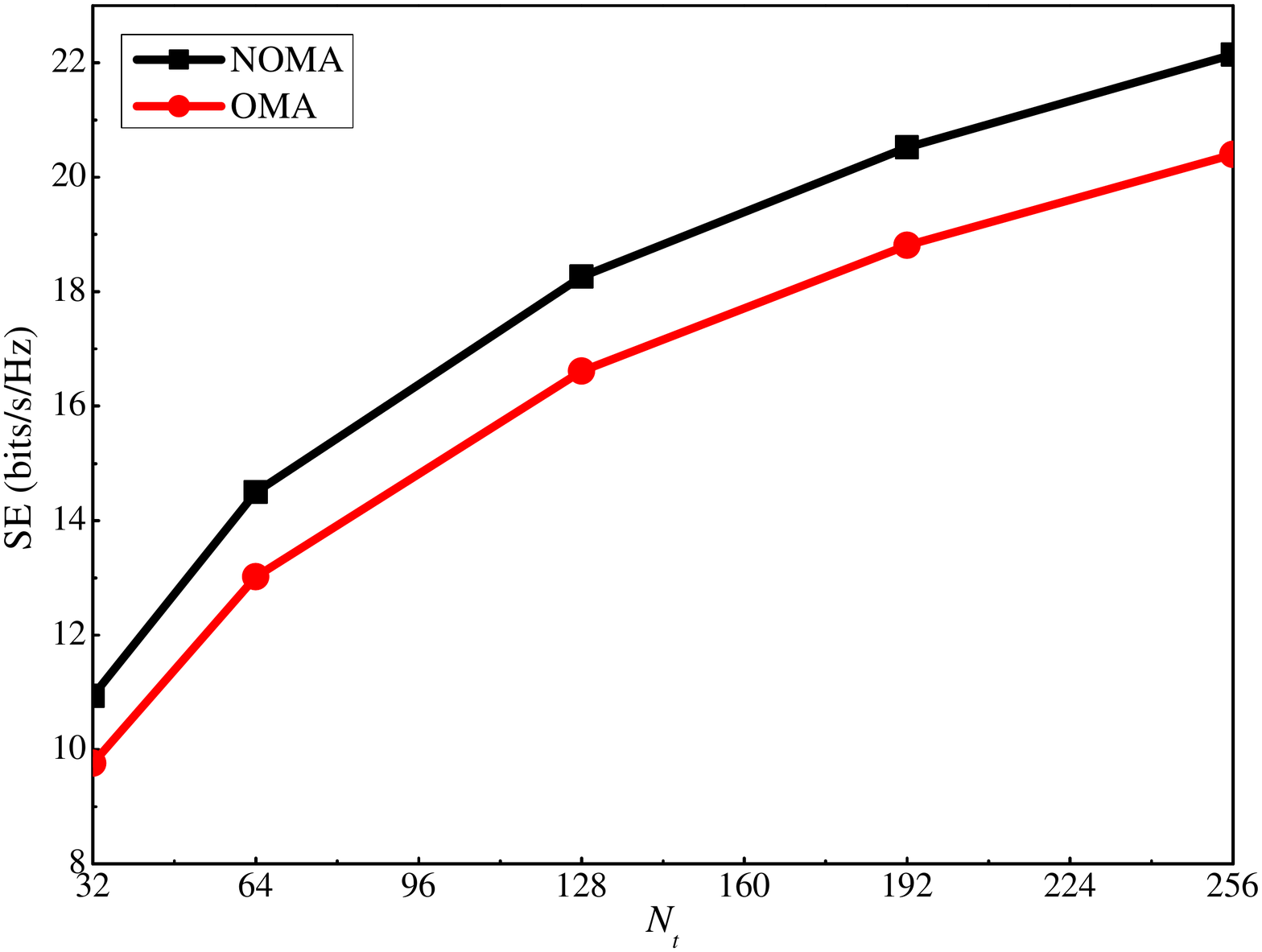}
  \caption{}
  \label{fig:sub1}
\end{subfigure}%
\begin{subfigure}{0.5\textwidth}
  \centering
  \includegraphics[width=1\linewidth]{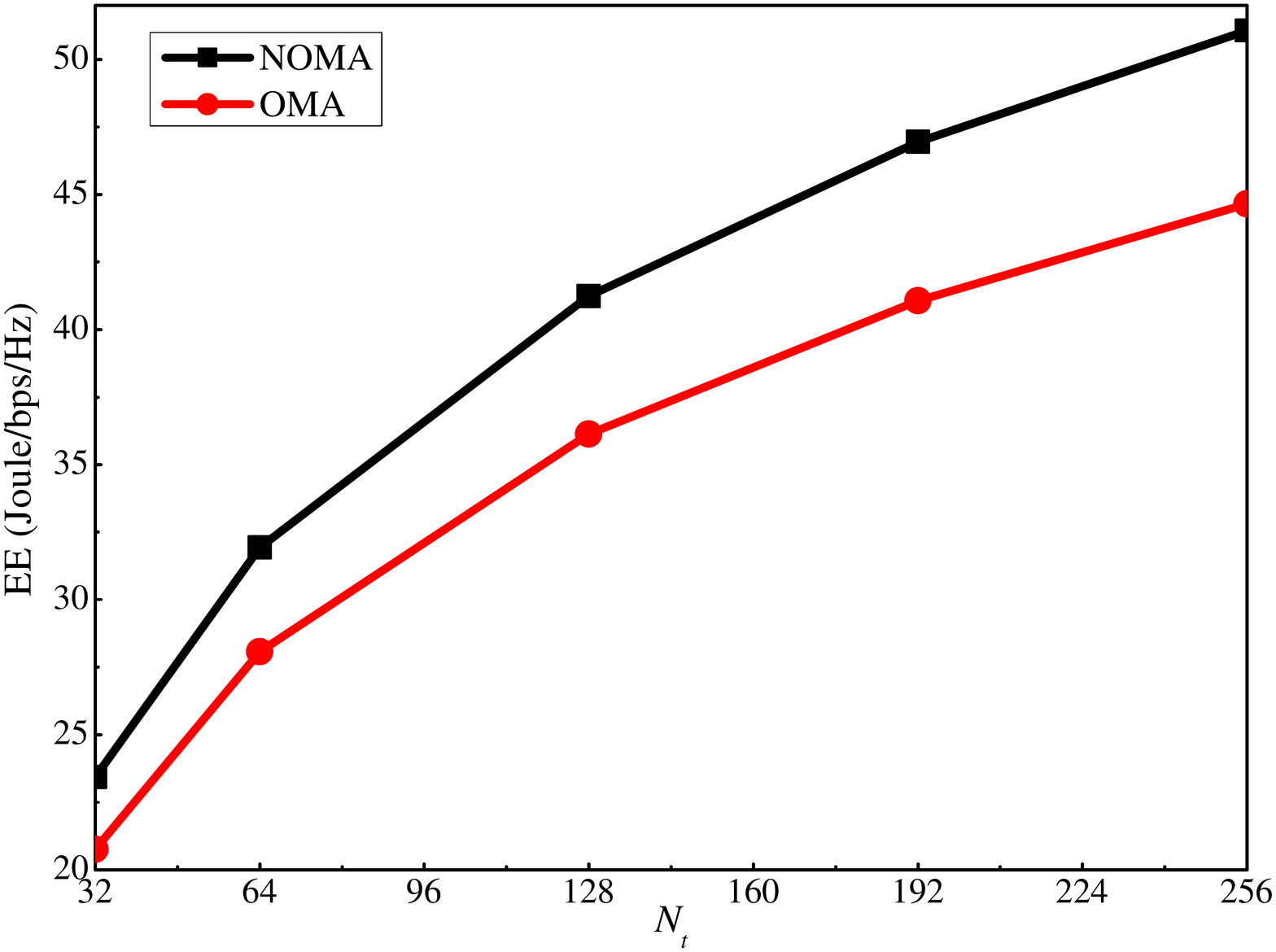}
  \caption{}
  \label{fig:sub2}
\end{subfigure}
\caption{{\color{black}Performance comparison for NOMA and OMA when the number of antenna varies: (a) SE; (b) EE.}}
\label{OMAvsNOMA}
\end{figure*}

Similar to the SE, we also show how the EE varies with the maximum uplink and downlink power in Figs.~\ref{EEvsPmax} and \ref{EEvsQmax}, respectively. For both cases, the EE first grows with the maximum power constraint, {\color{black}and} after a certain threshold, i.e., $P_{\rm{max}}=-20$ dB and $Q_{\rm{max}}=-10$ dB, {\color{black}it} remains unchanged even if the maximum power constraint continues to grow. This is because the slow increases in the SE cannot compensate for the power increment when the power is high, and thus, no more power will be consumed by the users to maximize the EE. By comparing the EE figures with the sum rate ones, i.e., Fig.~\ref{SEvsPUL} versus Fig.~\ref{EEvsPmax}, and Fig.~\ref{SEvsPDL} versus Fig.~\ref{EEvsQmax}, we can observe that the EE reaches the turning point at a smaller power value than the sum rate. This is because after the sum rate increment over the power declines to a certain value, no more extra power {\color{black}is} used to maximize the EE.

\begin{figure}
\centering
\includegraphics[width=0.65\textwidth]{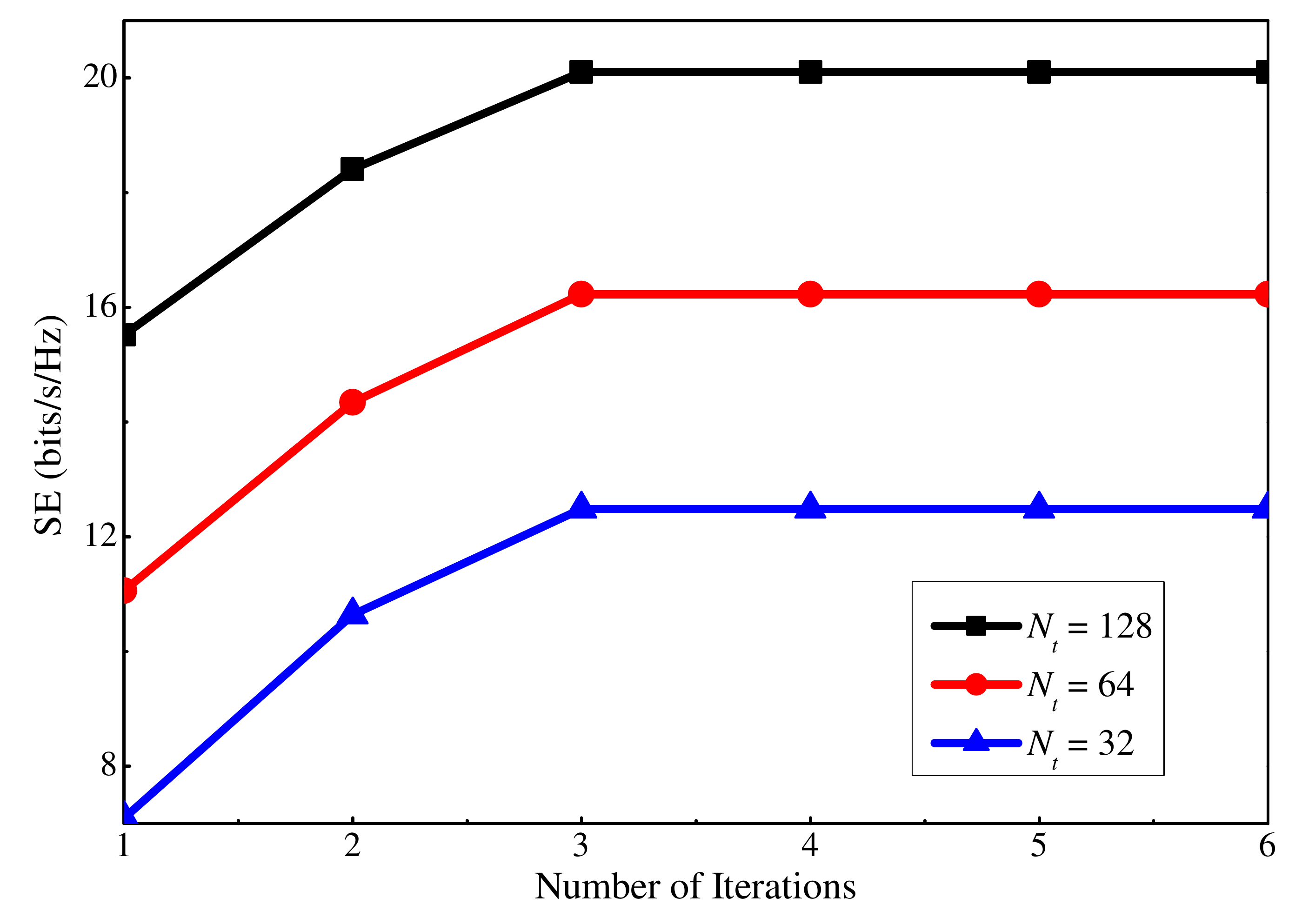}
\caption{{\color{black}Convergence of the proposed SE algorithm}.}
\label{SEvsIter}
\end{figure}

{\color{black}
The baseline massive MIMO-OMA can be considered as a special case of the proposed massive MIMO-NOMA scheme with just one user in each cluster. Accordingly, the legitimate achievable rate of the $m$-th user is:
\begin{equation}
	R^{OMA}_m = \left(1-\frac{\tau}{T}\right)\log_2\left(1+\frac{\kappa_m}{\sum\limits^3_{i=1} I_{m,i}+1}\right),
\end{equation} 
where
$\kappa_m = Q_m\beta_m\rho_m N_t$,
$I_{m,1} = Q_m\beta_m(1-\rho_m)$,
$I_{m,2} = \sum\limits_{i\neq m}^{M}Q_i\beta_m$, 
$I_{m,3} = Q_{m,0}\beta_m(1-\rho_m)+\sum\limits_{i\neq m}^{M}Q_{i,0}\beta_m$,
$Q_m$ is the downlink power for the $m$-th user,
$Q_{m,0}$ is the AN power for the $m$-th user,
$\beta_m$ is the large scale fading of the $m$-th user,
$\rho_m = \frac{P_m\beta_m\tau}{P_m\beta_m\tau+1}$,
$\tau$ is the length of training sequences that is the same as the NOMA case,
and $P_m$ is the uplink transmit power of the $m$-th user.
The achievable eavesdropping rate corresponding to the $m$-th user is:
\begin{equation}
R^{OMA}_{E,m} = 
\left(1-\frac{\tau}{T}\right)\log_2\left(1+
\frac{Q_m\beta_E}{\sum\limits_{i\neq m}^M Q_i\beta_E + \sum\limits_{i=1}^M Q_{i,0} \beta_E +1}
\right).
\end{equation}

The achievable secrecy rate of the $m$-th user is
\begin{equation}
R^{OMA}_{S,m} = [R^{OMA}_m-R^{OMA}_{E,m}]^+.
\end{equation}

In simulations, to compare the proposed massive MIMO-NOMA with the baseline massive MIMO-OMA, we consider a scenario with {four clusters and} two users in each cluster. TDMA is used for the baseline massive MIMO-OMA, and thus, each user in one cluster is only served half the time. Fig.~\ref{OMAvsNOMA} shows the corresponding SE and EE comparison between the considered schemes. It is clear that the proposed scheme outperforms the baseline massive MIMO-OMA when the number of antennas at the BS increases, which shows its superiority.

%
%
}

\begin{figure}
\centering
\includegraphics[width=0.65\textwidth]{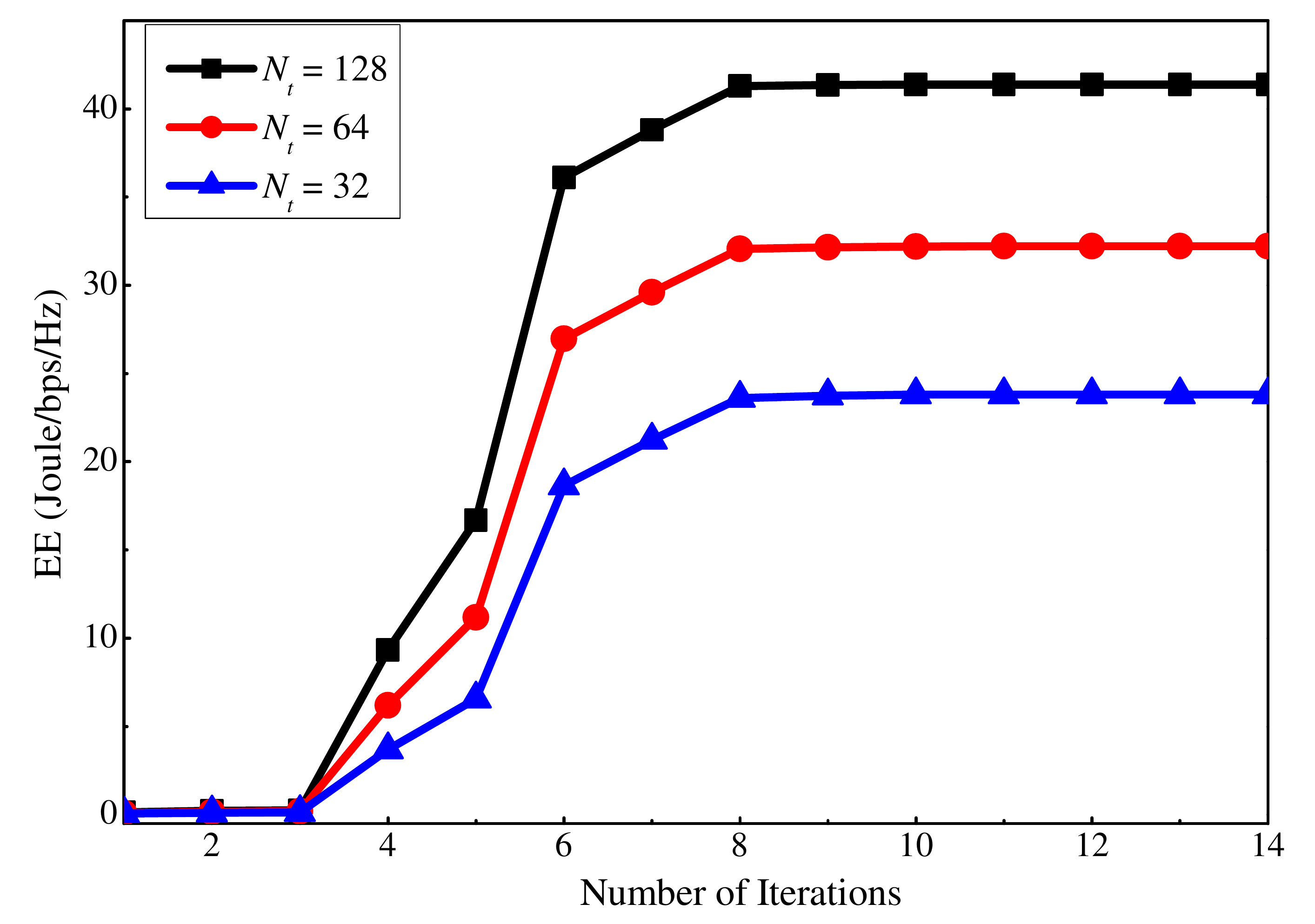}
\caption{{{\color{black}Convergence of the proposed EE algorithm}.}}
\label{EEvsIter}
\end{figure}

Finally, Figs.~\ref{SEvsIter} and \ref{EEvsIter} show how many iterations are required for the proposed SE and EE maximization algorithms to converge, respectively. Note that here an iteration means solving either the uplink or the downlink DC programming problem, which requires to solve an average of five convex problems according to the simulation. Results for three different antenna numbers are presented when $Q_{\rm{max}}=20$ dB and $P_{\rm{max}}=0$ dB. It can be seen that a small number of iterations are required for the proposed SE and EE maximization algorithms to converge.

\section{Conclusion}
In this paper, an {\color{black}AN-aided scheme has been proposed to ensure secrecy in massive MIMO-NOMA networks.
The} ergodic secrecy rate and its asymptotic value have been derived to spotlight the roles of key parameters on the secrecy performance of the considered system.
The results have revealed that {\color{black}with} a sufficiently large number of transmit antennas at the BS, only the illegitimate side is affected by the AN.
In addition, when the transmit power at the BS is high, the secrecy performance of a user is independent of the inter-cluster interference and AN and is determined by the uplink training process, which depends on the number of users in a cluster, the uplink transmit power, and the large-scale fading.
Besides, the results also suggest to keep the number of users in a cluster small for a better secrecy performance at each user and cluster.
Furthermore, numerical results validate that our proposed optimization algorithms can obtain significant improvements over {\color{black}the baseline algorithms}, i.e., Uplink PA, Downlink PA and Fixed PA, in terms of the sum ergodic secrecy rate and energy efficiency. This fully reveals the necessity of performing power optimization for the considered system, and the effectiveness of the proposed algorithms.
{\color{black}Finally, from the perspective of sum ergodic secrecy  rate and its energy efficiency, our proposed system surpasses the conventional massive MIMO-OMA system.}

\bibliographystyle{IEEEtran}
\balance
\bibliography{IEEEabrv,reference,mybibfile}

\end{document}